\documentclass[preprint,12pt]{elsarticle}

\usepackage{graphicx,color}
\usepackage{amssymb,amssymb,cleveref}

\journal{Computer Physics Communications}
\newcommand{\ud}{\mathrm{d}}
\newcommand{\ic}{\mathrm{i}}

\newcommand{\Ai}{\mathrm{ Ai}}
\newcommand{\Bi}{\mathrm{ Bi}}
\newcommand{\sgn}{\mathrm{sgn}}

\usepackage{soul}

\begin{document}

\begin{frontmatter}

%% Title, authors and addresses

%% use the tnoteref command within \title for footnotes;
%% use the tnotetext command for the associated footnote;
%% use the fnref command within \author or \address for footnotes;
%% use the fntext command for the associated footnote;
%% use the corref command within \author for corresponding author footnotes;
%% use the cortext command for the associated footnote;
%% use the ead command for the email address,
%% and the form \ead[url] for the home page:
%%
%% \title{Title\tnoteref{label1}}
%% \tnotetext[label1]{}
%% \author{Name\corref{cor1}\fnref{label2}}
%% \ead{email address}
%% \ead[url]{home page}
%% \fntext[label2]{}
%% \cortext[cor1]{}
%% \address{Address\fnref{label3}}
%% \fntext[label3]{}

\title{WKB approach to evaluate series of Mathieu functions in scattering problems}

\author[a]{Maxime Hubert}
\author[a]{R\'emy Dubertrand\corref{author}}

\cortext[author] {Corresponding author.\\\textit{E-mail address:} remy.dubertrand@ulg.ac.be}
\address[a]{ D\'epartement de Physique, CESAM, University of Li\`ege, 4000 Li\`ege,
Belgium}

\begin{abstract}
%% Text of abstract
The scattering of a wave obeying Helmholtz equation by an elliptic obstacle can be described exactly using series of Mathieu functions. This situation is relevant in optics, quantum mechanics and fluid dynamics.
We focus on the case when the wavelength is comparable to the obstacle size, when the most standard approximations fail. The approximations of the radial (or modified) Mathieu functions using WKB method are shown to be especially efficient, in order to precisely evaluate series of such functions. It is illustrated with the numerical computation of the Green function when the wave is scattered by a single slit or a strip (ribbon).
\end{abstract}


\end{frontmatter}

\section{Introduction}
\label{intro} 

The scattering of a (quantum or electromagnetic) wave by a slit or a strip has a long history \cite{Schwarzschild,Sieger,strutt,MorseRubenstein,KarpRussek} and still motivates studies, see e.g.~\cite{Nye,Brooker}. It has become the starting point for anyone interested in exact results in diffraction theory. For small or moderate values of the ratio between the slit/strip width and the wavelength, it is more efficient for a computational perspective to write the solutions as series of Mathieu functions \cite{KarpRussek,Brooker,Stamnes}.

%%% Historical/classical books, analytical approaches
If one seeks information regarding information about Mathieu functions, the most standard textbook is McLachlan's book \cite{McLachlan}. A more schematic description of the theory of Mathieu functions can be found in \cite{bateman3}. A detailed derivation of analytical approximations for Mathieu functions is performed in \cite{sips}.
Formulae useful for standard applications (esp. solutions of the first kind) have been collected in \cite{DLMF}. A more rigorous approach, close to our WKB perspective, has been developed to build asymptotic expansions for both angular and radial Mathieu functions in \cite{Langer}, and more recently in \cite{Ogilvie}. A WKB approach for the angular functions has recently been developed in \cite{ODell}.
Finally, for a recent pedagogical introduction to Mathieu functions, we refer to \cite{Gutierrez}.
% As the standard formula are known are known to be inaccurate at small values of the argument power law expansion have been derived in \cite{Larsen2009}. 

%%% Numerical evaluation, libraries
Angular Mathieu functions can be easily found in most of numerical softwares (Maple, Mathematica, Matlab) and libraries (GSL, SciPy). We want to stress that this is different for radial Mathieu functions. Especially third kind radial Mathieu functions, which are crucial for scattering problems, can be only found in Matlab to our knowledge, see e.g.~\cite{Cojocaru}, and SciPy \cite{SciPy_web}. Subroutines available for radial Mathieu functions of all kinds can be seen in \cite{Zhang} and will be used in the following to estimate the accuracy of our method.

%%% Mathieu function in physical problems
It is usual to display the eigenmodes of an elliptic cavity as a check for the numerical computation of Mathieu functions \cite{gheorghiu}. We want to pursue a clearly distinct route and we will be focused on scattering problems. 
The first computation of scattered far field by a slit or a strip using numerical evaluation of Mathieu function was displayed in \cite{MorseRubenstein}, and a comparison between theory and experiments was performed in \cite{hsu}. 
It is also worth stressing that Mathieu functions can appear in other physical situations, especially for the scattering by a $1/r^4$ potential \cite{OMalley,Hunter}, with a recent revival in quantum defect theory for cold atom experiments\cite{Gao}.

In Section~\ref{definitions}, the definitions and the notations are set for our results. 
%Mathieu function can have slightly different normalisations, especially when
In Section~\ref{num_eval_Mathfunc}, the standard way to evaluate numerically Mathieu functions is recalled. 
In Section~\ref{WKB_Mathfunc}, we suggest a method inspired from semiclassical quantum mechanics, the WKB method, in order to have reliable formulae in this regime. In Section~\ref{Green}, the accuracy of our method is more precisely quantified and illustrated by computing the Green function for scattering problems.
In Section~\ref{conclusion}, some concluding remarks are listed. More technical details and further checks have been moved to Appendices.

\section{Definition \& notations for Mathieu functions}
\label{definitions}
For pedagogical purposes Mathieu functions are introduced as
the solution of  Mathieu equations. The notations for the related quantities are also carefully listed, as several notations/conventions can be adopted, see more details in \ref{notations}.

Mathieu functions appear very naturally when looking at the dynamics of a wave. Consider Helmholtz equation for a wave in the two-dimensional Euclidean plane:
\begin{equation}
  \label{wave_eq}
  (\Delta+k^2)\varphi=0\ ,
\end{equation}
with $k=2\pi/\lambda$ being the wave number and $\varphi$ the wave amplitude.
Eq.~(\ref{wave_eq}) plays a key role in many areas of physics. It is the stationary form for an electromagnetic wave in the vacuum and for a free quantum particle. One can introduce the elliptic coordinates $(u,v)$:
\begin{eqnarray}
 x&=& \frac{a}{2} \cosh u\cos v\ ,\label{eqcoordell1}\\
y&=&\frac{a}{2} \sinh u\sin v\ ,  u\ge 0,\ -\pi< v\le \pi\label{eqcoordell2}
\end{eqnarray}
where $(x,y$) are 2D Cartesian coordinates, and $a$ is the distance between the two foci of the coordinate system. 
One can easily check that the curves with constant $u$ are confocal ellipses whereas curves with constant $v$ are the branches of confocal hyperbolae.
Using the expression of the Laplacian in elliptic coordinates:
\begin{equation}
    \Delta \varphi(u,v)=\frac{2}{\left(\frac{a}{2}\right)^2\left(\cosh 2u-\cos 2v\right)}\left(\frac{\partial^2 }{\partial u^2} + \frac{\partial^2 }{\partial v^2} \right)\varphi(u,v)\ ,
\end{equation}
it is possible to write a separable solution of Eq.~(\ref{wave_eq}), by defining $\varphi(u,v)=U(u)V(v)$ under the conditions:
\begin{eqnarray}
  \label{eqmathieu}
  V''(v)+\left[h-2\theta\cos(2v)\right]V(v)=0
  &\ ,\\
  \label{eqmathieum}
  U''(u)-\left[h-2\theta\cosh(2u)\right]U(u)=0
  &\ .
\end{eqnarray}
where we introduced the parameter $\theta=\left(\frac{ka}{4}\right)^2$, and $h$ is a numerical separation constant.
The main goal of this study is the accurate numerical evaluation of 
the solutions of both Eqs.~(\ref{eqmathieu}) and (\ref{eqmathieum}) with a focus on the regime $\theta\sim 1$.

\subsection{Angular Mathieu equation}

From Floquet theory Eq.~(\ref{eqmathieu}) admits periodic solutions for a \emph{discrete} set of values of $h(\theta)$, called the characteristic value. For a fixed $\theta$ and $h=h(\theta)$ the periodic solution of Eq.~(\ref{eqmathieu}) --- unique up to normalization, can be made real for real $v$; it is referred to as Mathieu function. 
It is common to distinguish between two sets of characteristic curves in the $(\theta,h)$ plane, which are associated to periodic solutions. They differ by the symmetry of the associated solution of Eq.~(\ref{eqmathieu}).

If one is looking for a periodic even solution of Eq.~(\ref{eqmathieu}), the characteristic value will denoted by $a_n(\theta)$. The corresponding angular Mathieu function is $ce_n(\theta,v)$. Note that $n$ is here a non negative integer: $n\ge 0$.

If one is looking for a periodic odd solution of Eq.~(\ref{eqmathieu}), the characteristic value will denoted by $b_n(\theta)$. The corresponding angular Mathieu function is $se_n(\theta,v)$. Note that $n$ has become a positive integer: $n>0$.

There has been a long (and still today) standing issue about the normalization of Mathieu function. We will use the one from the oldest papers, which first described analytical formulae.
Mathieu functions are normalized so as to form an orthogonal family:
\begin{equation}
\label{orthoM}
  \int_0^{2\pi} ce_n(\theta,v) ce_m(\theta,v)\ud v =\int_0^{2\pi} se_n(\theta,v) se_m(\theta,v)\ud v =\pi \delta_{m,n}\ ,
\end{equation}
where $\delta_{m,n}$ denotes Kronecker symbol. A last condition is required to uniquely define Mathieu functions:
\begin{equation}
  ce_n(\theta,0)>0,\quad \frac{\partial }{\partial v}se_n(\theta,v)\Big|_{v=0}>0\ .\label{condA2n}
\end{equation}

\subsection{Radial Mathieu equation}

Following \cite{bateman3}, we will denote by $Ce_n(\theta,u)$ and $Se_n(\theta,u)$ the radial Mathieu functions of the first kind associated to two different values of $h$: $h=a_n(\theta)$ and $h=b_n(\theta)$ respectively. Note that these functions are real valued when $u$ is real and $\theta$ is real positive.
This notation is also intuitive as it stresses the simple relations between these solutions and the angular Mathieu functions:
\begin{equation}
  Ce_n(\theta,u)=ce_n(\theta,\ic u),\quad Se_n(\theta,u)=-\ic se_n(\theta,\ic u)\ .\label{Ang_to_rad}
\end{equation}
In analogy with the theory of Bessel equation, it is convenient to define another solution of (\ref{eqmathieum}). We denote by $Fey_n(\theta,u)$, resp. $Gey_n(\theta,u)$, the radial Mathieu functions of the second kind associated to $a_n(\theta)$, resp.  $b_n(\theta)$, which are also real valued in the above mentioned range of $u$ and $\theta$. They can be shown to be linearly independent from the radial Mathieu function of the first kind for both values of $h$.
In the rest of the paper, and in connection with applications in diffraction theory, we define the radial Mathieu functions of the third kind:
\begin{eqnarray}
  \label{defMe}
  Me^{(1)}_n(\theta,u)&=&Ce_n(\theta,u)+\ic Fey_n(\theta,u),\nonumber\\
 Me^{(2)}_n(\theta,u)&=&Ce_n(\theta,u)-\ic Fey_n(\theta,u)\nonumber\\ % h=a_n(\theta)\\
  \label{defNe}
  Ne^{(1)}_n(\theta,u)&=&Se_n(\theta,u)+\ic Gey_n(\theta,u),\nonumber \\
 Ne^{(2)}_n(\theta,u)&=&Se_n(\theta,u)-\ic Gey_n(\theta,u) \nonumber  %& h=b_n(\theta)\nonumber
\end{eqnarray}
As a conclusion two linearly independent solutions of (\ref{eqmathieum}) can be chosen as follows:
\begin{itemize}
\item $Ce_n(\theta,u)$ and $Me^{(1)}_n(\theta,u)$ for $h=a_n(\theta)$,
\item $Se_n(\theta,u)$ and $Ne^{(1)}_n(\theta,u)$ for $h=b_n(\theta)$.
\end{itemize}
They can be shown to be always linearly independent for real positive $\theta$, e.g. by looking at their asymptotics for large $u$.
%It can be shown \cite{McLachlan} that any radial Mathieu function can be expanded as series of Bessel functions. 

\section{Numerical evaluation of Mathieu functions}
\label{num_eval_Mathfunc}

It is worth highlighting here that we aim at an accurate, fast and light numerical method, which can be applied to scattering problems by an obstacle. 
We also stress that the formulae below are primarily valid for moderate values of $\theta$ and should remain accurate when going to the large $\theta$ regime.
In order to make the formulae more compact, the explicit dependence on $\theta$ will be dropped throughout this Section.

\subsection{Angular Mathieu functions}

As they are periodic, angular Mathieu functions can be expanded into Fourier series:
\begin{eqnarray}
  \label{Fourier_ce_even}
  ce_{2n}(v)&=&\displaystyle\sum_{p=0}^\infty A^{(2n)}_{2p}\cos(2 p v) ,\\
  \label{Fourier_ce_odd}
  ce_{2n+1}(v)&=&\displaystyle\sum_{p=0}^\infty A^{(2n+1)}_{2p+1}\cos\left[(2p+1)v\right], \\
  \label{Fourier_se_even}
  se_{2n}(v)&=&\displaystyle\sum_{p=1}^\infty B^{(2n)}_{2p} \sin(2 p v) ,\\
  \label{Fourier_se_odd}
  se_{2n+1}(v)&=&\displaystyle\sum_{p=0}^\infty B^{(2n+1)}_{2p+1} \sin\left[(2p+1)v\right]\ .
\end{eqnarray}
In particular one can notice that Mathieu functions of even index are $\pi-$periodic whereas those of odd index are $2\pi-$periodic.
For the sake of exhaustiveness, our normalization of Mathieu function, see (\ref{orthoM}), can be written:
\begin{eqnarray}
  2A^{(n)\; 2}_{0}+ \sum_{p=1}^\infty A^{(n)\; 2}_{p}= 1,& \displaystyle\sum_{p=0}^\infty A^{(n)}_{p} > 0\ , \label{normAn}\\
  \sum_{p=1}^\infty B^{(n)\; 2}_{p}= 1,& \displaystyle\sum_{p=1}^\infty p B^{(n)}_{p} > 0\ .\label{normBn}
\end{eqnarray}

\begin{figure}[!ht]
  \begin{center}
    \includegraphics[width=\linewidth]{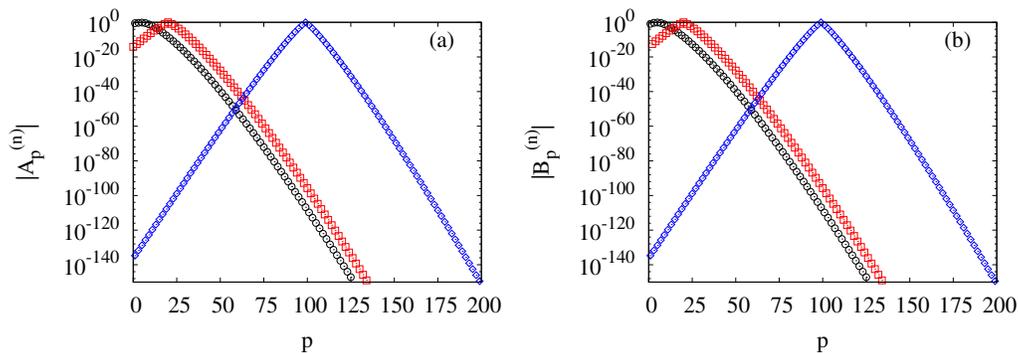}
  \end{center}  
  \caption{Modulus of Fourier coefficients of angular Mathieu functions of index $n$ in semi log scale. 
Black circles: $n=5$. Red squares: $n=20$. Blue diamonds: $n=99$. (a): even symmetry class. (b): odd symmetry class.}
  \label{example_AnBn}
\end{figure}

%The angular Mathieu functions are evaluated using their Fourier expansions, see (\ref{Fourier_ce_even}), (\ref{Fourier_ce_odd}),(\ref{Fourier_se_even}) and (\ref{Fourier_se_odd}). This simple way 
The Fourier series expansion 
is also a very efficient procedure to evaluate the function numerically as the Fourier coefficients $A^{(n)}_{p}$ and $B^{(n)}_{p}$ decreases rapidly (typically exponentially) for $p \gg n$. We need some coefficients with very high accuracy in the following, which we could obtain via a standard eigenvalue method. The details of the method can be found in \ref{Fourier_coeff}. The results are illustrated in Fig.~\ref{example_AnBn}. Notice especially the very small values for the first nonzero Fourier coefficient when $n$ increases.

\subsection{Standard evaluation of radial Mathieu functions}

The radial Mathieu functions are more challenging to evaluate numerically.
The most natural and usual way, which is reminded below, is to write them as a series of product of Bessel functions, see e.g.~\cite{McLachlan}. Each term is proportional to the Fourier coefficient of the corresponding angular Mathieu function. Hence the obtained series for each radial Mathieu function comprises terms with large variations and alternating signs.
Further it is recalled that we will need to compute infinite series of Mathieu functions to solve our physical problem.

Below are given the expansion for each
third kind radial Mathieu functions, first for the even symmetry class:
\begin{equation}
  \label{Me2n_seriesfin}
  Me_{2n}^{(1)}(u)=C_{2n}%\frac{ce_{2n}(\pi/2) ce_{2n}(0)}{ A^{(2n)\; 2}_{0}}
\sum_{p=0}^\infty (-1)^p A^{(2n)}_{2p} H_p^{(1)}\left(\sqrt{\theta}e^u\right) J_p\left(\sqrt{\theta}e^{-u}\right)\ ,
\end{equation}
\begin{eqnarray}
 Me_{2n+1}^{(1)}(u)=&C_{2n+1}%-\frac{ce'_{2n+1}(\pi/2) ce_{2n+1}(0)}{ \sqrt{\theta}A^{(2n+1)\; 2}_{1}}
\displaystyle\sum_{p=0}^\infty (-1)^p A^{(2n+1)}_{2p+1} \times\nonumber\\
&\left[ H_{p+1}^{(1)}\left(\sqrt{\theta}e^u\right) J_p\left(\sqrt{\theta}e^{-u}\right)+H_{p}^{(1)}\left(\sqrt{\theta}e^u\right) J_{p+1}\left(\sqrt{\theta}e^{-u}\right)\right]\ ,
  \label{Me2n+1_seriesfin}
\end{eqnarray}
where $A^{(n)}_{p}$ %is the Fourier coefficient of the angular Mathieu function 
is defined in (\ref{Fourier_ce_even}) and (\ref{Fourier_ce_odd}), 
and second for the odd symmetry class:
\begin{eqnarray}
Ne_{2n+1}^{(1)}(u)=&D_{2n+1}
\displaystyle\sum_{p=0}^\infty (-1)^p B^{(2n+1)}_{2p+1}\times\nonumber\\
& \left[ H_{p+1}^{(1)}\left(\sqrt{\theta}e^u\right) J_p\left(\sqrt{\theta}e^{-u}\right)-H_{p}^{(1)}\left(\sqrt{\theta}e^u\right) J_{p+1}\left(\sqrt{\theta}e^{-u}\right)\right]\ ,
  \label{Ne2n+1_seriesfin}
\end{eqnarray}
\begin{eqnarray}
  \label{Ne2n_seriesfin}
  Ne_{2n}^{(1)}(u)=&D_{2n}
\displaystyle\sum_{p=1}^\infty (-1)^p B^{(2n)}_{2p} \times\nonumber\\
&\left[ H_{p-1}^{(1)}\left(\sqrt{\theta}e^u\right) J_{p+1}\left(\sqrt{\theta}e^{-u}\right)-H_{p+1}^{(1)}\left(\sqrt{\theta}e^u\right) J_{p-1}\left(\sqrt{\theta}e^{-u}\right)\right]\ ,
\end{eqnarray}
where $B^{(n)}_{p}$ is %now stands for the Fourier coefficient of the angular Mathieu function 
defined in (\ref{Fourier_se_even}) and (\ref{Fourier_se_odd}). 
The prefactors in these expansions are following our definition of the third kind functions \cite{McLachlan,sips}:
\begin{eqnarray}
  C_{2n}&=\frac{\pi Me^{(1)\;\prime}_{2n}(0)}{2\ic\ ce_{2n}(\pi/2)}&=\frac{ce_{2n}(\pi/2) ce_{2n}(0)}{ A^{(2n)\; 2}_{0}}\ ,
\label{C2n}\\
  C_{2n+1}&=-\frac{\pi \sqrt{\theta} Me^{(1)\;\prime}_{2n+1}(0)}{2\ic\ ce'_{2n+1}(\pi/2)}&=-\frac{ce'_{2n+1}(\pi/2) ce_{2n+1}(0)}{ \sqrt{\theta}A^{(2n+1)\; 2}_{1}} \ , \label{C2n+1}\\
  D_{2n+1}&= -\frac{\pi \sqrt{\theta} Ne^{(1)}_{2n+1}(0)}{2\ic\ se_{2n+1}(\pi/2)} &= \frac{se_{2n+1}(\pi/2) se'_{2n+1}(0)}{ \sqrt{\theta}B^{(2n+1)\; 2}_{1}} \ ,\label{D2n}\\
  D_{2n}&=\frac{\pi \theta Ne^{(1)}_{2n}(0)}{2\ic\ se'_{2n}(\pi/2)} &=-\frac{se'_{2n}(\pi/2) se'_{2n}(0)}{\theta B^{(2n)\; 2}_{2}}\label{D2n+1}\ .
\end{eqnarray}
While the above expansions are exact, and can be used in the small $\theta$ regime they become rapidly inaccurate for $n\sim 10$ and $\theta\sim 1$, especially at small values of $u$. The main reason is that the prefactors of the expansions increase more than exponentially with $n$, see Fig.~\ref{example_AnBn}, so that numerical instabilities are magnified.
%Examples of the poor convergence of the series (\ref{Me2n+1_seriesfin}) are illustrated in Fig.~\ref{series_vs_WKB}.

\section{WKB approach to evaluate the radial Mathieu functions}
\label{WKB_Mathfunc}

In this Section we detail how to obtain accurate approximations of radial Mathieu functions, which are especially efficient to use in a series expansion. The main justification of our approach is to notice that these functions can be seen as the eigenfunctions of a quantum particle trapped in an exponentially deep well in one dimension. 

\subsection{Reminder of WKB method}
\label{crashWKB}

Consider a quantum particle of mass $m$ moving along the half line $x\ge 0$, subject to the potential $V(x)$. In the following only stationary states will be considered. Following standard non relativistic quantum mechanics \cite{Landau} the particle is described by a wave function $\psi(x)$, which obeys the stationary Schr\"odinger equation:
\begin{equation}
  \label{stat_schro}
  % -\frac{\partial^2}{\partial x^2} \psi(x) -\frac{2mV_0}{\hbar^2} \cosh(2x)\psi(x)= -\frac{2m E}{\hbar^2} \psi(x)\ .
 \frac{\partial^2}{\partial x^2} \psi(x) +\frac{2m}{\hbar^2}\left[E-V(x)\right]\psi(x)=0\ .
\end{equation}
where $\hbar$ denotes the Planck constant. 
The WKB method, sometimes denoted by eikonal method, is a powerful technique, which allows to write an uniform approximation of the solutions of Eq.~(\ref{stat_schro}) in the regime when $\hbar$, considered as a varying parameter, is small \cite{Landau}. %In that regime one can write explicit formulae to approximate the solutions. 
This approximation is written as asymptotic series, and we will be restricted here to the leading term of this expansion. The next sub-leading term can be found e.g. chap.~$7$ in \cite{Landau}.
Three cases need to be distinguished depending on whether the prefactor $E-V(x)$ remains positive, negative or has a zero. Below it will be assumed that $x$ belongs to the range $[x_i;x_f]$

The first case to be treated is when $E-V(x)$ stays negative along the considered range in $x$. In the corresponding classical problem it means that the classical particle is trapped inside the potential well $V(x)$. Introduce the classical action:
\begin{equation}
  \label{Sclass}
   S^{(1)}(x)=\int_{x_i}^x \sqrt{2m\left[E-V(t) \right]}\ud t\ ,
\end{equation}
and the classical momentum:
\begin{equation}
  \label{pclass}
  p^{(1)}(x)=S^{(1)\;\prime}(x)=\sqrt{2m\left[E-V(x) \right]}\ ,
\end{equation}
where the prime denotes differentiation with respect to $x$.
The WKB approximation of the solution of (\ref{stat_schro}) in this situation is, see e.g.~\cite{Landau}:
\begin{equation}
  \label{WKB_inside}
  \psi_{{\rm WKB}}^{(1)}(x)=\frac{A_+ \;e^{S^{(1)}(x)/\hbar}+A_-\;e^{- S^{(1)}(x)/\hbar}}{|p^{(1)}(x)|^{1/2}}\ ,
\end{equation}
 $A_+$ and $A_-$ are constants, which are fixed by the conditions at the ends of the range $x=x_i$ , and $x=x_f$. It is easy to check that the ansatz (\ref{WKB_inside}) obeys the equation (\ref{stat_schro}) with an error of order ${\cal O}(1)$ in the regime $\hbar\to 0$. 

The second case is when $E-V(x)$ stays positive for $x$ in $[x_i;x_f]$. The classical action and momentum are now defined as:
\begin{equation}
  \label{Sclass2}
   S^{(2)}(x)=\int_{x_i}^x \sqrt{2m\left[V(t)-E \right]}\ud t\ ,\quad p^{(2)}(x)=S^{(2)\;\prime}(x)=\sqrt{2m\left[V(x)-E \right]}\ .
\end{equation}
With those new definitions this case can be treated in a similar way as the previous one, and the WKB ansatz is:
\begin{equation}
  \label{WKB_outside}
  \psi_{{\rm WKB}}^{(2)}(x)=\frac{B_+ \;e^{\ic S^{(2)}(x)/\hbar}+B_-\;e^{-\ic  S^{(2)}(x)/\hbar}}{|p^{(2)}(x)|^{1/2}}\ ,
\end{equation}
where $B_+$, and $B_-$ are constants fixed by the boundary condition.

The third case is slightly more complicated as it amounts to approximate the solution of (\ref{stat_schro}) around a point $x_*$ defined by 
\begin{equation}
  \label{def_turning_point}
  V(x_*)=E\ .
\end{equation}
Such a point is usually called a turning point. It plays a special role, especially because, as mentioned above, the error term in the WKB expansion diverges. It is also important in order to ``connect'' the above WKB solutions when crossing a turning point.
It will be assumed here that the first derivative of the potential does not vanish at the turning point. It is required to redefine the action and the momentum:
\begin{equation}
  \label{Sclass3}
   S^{(3)}(x)-S_*=\int_{x_*}^x \sqrt{2m\big| E-V(t)\big|}\ud t\ ,
\quad p^{(3)}(x)=S^{(3)\;\prime}(x)=\sqrt{2m\big| E-V(x)\big| }\ ,
\end{equation}
with
\begin{equation}
  \label{defSst}
  S_*=\int_{x_i}^{x_*} \sqrt{2m\big| E-V(t)\big|}\ud t
\end{equation}

Then the WKB ansatz is, see e.g.~\cite{Langer}:
\begin{eqnarray}
  \psi_{{\rm WKB}}^{(3)}(x)=& |p^{(3)}(x)|^{-\frac{1}{2}}\left( \frac{|S^{(3)}(x)-S_*|}{\hbar}\right)^{\frac{1}{6}} \Bigg\{C_+\; \Ai\left[\pm \sgn(x-x_*) \left(\frac{3}{2}\frac{|S^{(3)}(x)-S_*|}{\hbar}\right)^{\frac{2}{3}} \right]\nonumber\\ & +C_-\; \Bi\left[ \pm \sgn(x-x_*)\left(\frac{3}{2}\frac{|S^{(3)}(x)-S_* |}{\hbar}\right)^{\frac{2}{3}} \right] \Bigg\} \ ,
\label{WKB_turning}
\end{eqnarray}
where $C_+$, and $C_-$ are constants fixed by the boundary condition. $\Ai(x)$, and $\Bi(x)$ denote the two linearly independent solutions of the Airy equation %We also defined $S^{(3)}_*=S^{(3)}(x_*)$. 
%We introduced both solutions of the Airy equation:
\begin{equation}
  y''(x)-x y(x)=0\ .
\end{equation}
The remaining signs in (\ref{WKB_turning}) depend on the direction across which the turning point is visited.
The formulae (\ref{WKB_inside}), (\ref{WKB_outside}) and (\ref{WKB_turning}) allow one to write a semiclassical approximation of a solution to Schr\"odinger equation in a longer range, connecting suitable expressions at each turning point.
In prevision of the future numerical method applied to Mathieu functions, we illustrate the result of the WKB approach in the following example ($x\ge 0$):
\begin{equation}
  \label{WKB_ex}
  V(x)=-V_0 \cosh({2x}), \quad V_0>0\ .
\end{equation}
We seek approximate solutions of (\ref{stat_schro}) in the regime $E<0$. {Until the end of this Section we drop unnecessary constants and take $m=1/2$ and $\hbar=1$.}
For the potential (\ref{WKB_ex}) along the half line, there is only one turning point defined by (\ref{def_turning_point}) along the half line. For $0<x<x_*$, we can use the above formula (\ref{WKB_inside}). As mentioned above the arbitrary coefficients are fixed from the boundary condition. Here Neumann boundary condition is imposed at $x=0$ such that $A_+=A_-$ in (\ref{WKB_inside}). Therefore the WKB approximation for the solution of the Schr\"odinger equation is for $0<x<x_*$:
\begin{equation}
  \label{WKB_ex_in}
  \psi_{}^{(1)}(x)=A \frac{\cosh\left({S^{(1)}(x)}\right)}{|S^{(1)\;\prime}(x)|^{1/2}}\ ,\quad
S^{(1)}(x)=\int_0^x {\sqrt{ -V_0\cosh(2t) -E}}\ud t\ ,
\end{equation}
with a constant prefactor $A$, which is determined at the very end by a normalization constraint.

For $x\simeq x_*$ the formula (\ref{WKB_turning}) should be used. Notice that, as $x$ increases, the wave function changes from (\ref{WKB_inside}) to (\ref{WKB_outside}) so that the minus sign has to be chosen in (\ref{WKB_turning}). The constant coefficients are fixed in order to fulfill the continuity of both the function and its first derivative. 
%As explained in \ref{connectionWKB} this leads to:
For the sake of brevity we only give the result:
\begin{eqnarray}
    \psi_{}^{(3)}(x)= B \frac{|S^{(3)}(x)-S_*|^{\frac{1}{6}}}{|S^{(3)\;\prime}(x)|^{\frac{1}{2}}}\Bigg\{& e^{S_*}\Ai\left[\sgn(x_*-x) \left(\frac{3}{2}|S^{(3)}(x)-S_*|\right)^{\frac{2}{3}} \right]+\nonumber\\
&  \frac{e^{-S_*}}{2} \Bi\left[\sgn(x_*-x)\left(\frac{3}{2}|S^{(3)}(x)-S_* |\right)^{\frac{2}{3}}\right]\Bigg\} \label{WKB_ex_turn}
\end{eqnarray}
with
\begin{equation}
  \label{}
  S^{(3)}(x)=\int_0^x \sqrt{\left|E +V_0\cosh(2t) \right|}\ud t\ .
\end{equation}
The constant prefactor is related to the previously introduced one via
\begin{eqnarray}
  B=A\sqrt{\pi}\left(\frac{3}{2}\right)^{1/6}\ .\nonumber
\end{eqnarray}

At last for $x>x_*$ the formula (\ref{WKB_outside}) is appropriate. Again the coefficients are fixed in order to ensure continuity.
This leads to:
\begin{eqnarray}
    \psi_{}^{(2)}(x)= 2A\frac{e^{S_*}}{S^{(2)\;\prime}(x)^{1/2}} \cos\left[ S^{(2)}(x)-S_*-\frac{\pi}{4} \right] \label{WKB_ex_out}
\end{eqnarray}
with
\begin{equation}
  \label{}
  S^{(2)}(x)-S_*=\int_{x_*}^x \sqrt{E +V_0\cosh(2t) }\ud t,\ S^{(2)\;\prime}(x)=\sqrt{E +V_0\cosh(2x) }\ .
\end{equation}
As mentioned above, there is eventually one overall constant factor, $A$, which can be fixed by normalization or by prescribing a fixed value at $x=0$.

\begin{figure}[!ht]
  \begin{center}
    \includegraphics[width=.6\textwidth]{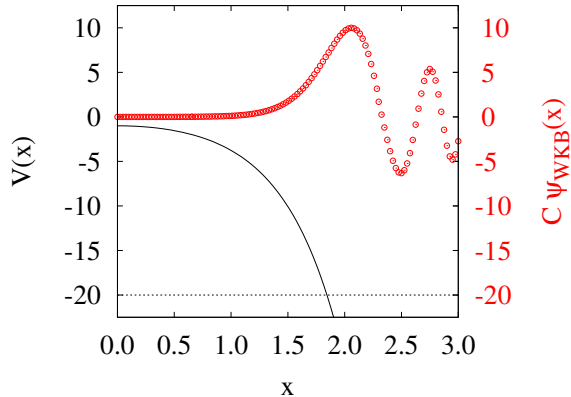}
  \end{center}  
  \caption{WKB approximation for the solution of (\ref{stat_schro}) when $V(x)=-V_0\cosh(x)$ and $E<0$. The black full curve stands for the potential (\ref{WKB_ex}) with $V_0=1$. The black dashed curve indicates the energy, here $E=-20$ hence $x_*\sim {1.84}$. The red circles correspond to the solution of (\ref{stat_schro}) associated to this value of the energy within the WKB approximation. The solution (\ref{WKB_ex_in}) is computed for $x$ between $0$ and {$0.66$}, the formula (\ref{WKB_ex_turn}) is used for $x$ between {$0.66$ and $2.3$}. Last the expression (\ref{WKB_ex_out}) is used for {$x>2.3$}. For sake of illustration the wave function has been dilated by a constant factor $C$.
  }
  \label{WKB_diag}
\end{figure}

An illustration of such an obtained approximation is shown in Fig.~\ref{WKB_diag}.

\subsection{Application to Mathieu equation}

If the following substitutions are made in (\ref{stat_schro}):
\begin{equation}
  \label{subs}
 V(x)=-V_0 \cosh(2x),\quad \theta=2 m V_0/\hbar^2,\quad h=-2m E/\hbar^2\ ,
\end{equation}
one immediately recovers (\ref{eqmathieum}).
In other words the radial Mathieu functions can be seen as the eigenstates of the quantum scattering problem by the potential (\ref{WKB_ex}) when the associated energy is negative\footnote{such that $h$ is positive as it is always assumed throughout our paper.}. Note that the scattering problem admits eigenstates for every real values of $E$. The separation of variables as described in Sect.~\ref{intro}, leads us to focus on special values of the energy level in the quantum perspective. 
 
Following the previous Section, the exact solution of such scattering problem can be efficiently approximated by explicit formulae in the semiclassical regime, i.e. when $h$ is large. This is especially fulfilled when $n$ is large as both $a_n$ and $b_n$ grows like $n^2$ when $n$ goes to infinity \cite{McLachlan}.

\subsubsection{WKB formulae for the even Mathieu functions}

Following the reminder in Sect.~\ref{crashWKB}, a very important quantity is the action associated to the problem (\ref{stat_schro}):
\begin{equation}
  \label{defS}
  S(u)=\int_0^u\sqrt{a_{n}-2\theta\cosh(2t)}\ud t\textrm{ and } S'(u)=\sqrt{a_{n}-2\theta\cosh(2 u)}\ ,
\end{equation}
where the prime denotes the derivative with respect to $u$.
%Notice the modulus to account for the contribution on both sides of the potential barrier, i.e. in the classically allowed and in the classically forbidden regions.

When $V(x)<a_n$, the action (\ref{defS}) is real, and rewriting (\ref{WKB_inside}) using (\ref{WKB_ex}) leads to the following approximations for the Mathieu functions:
\begin{eqnarray}
  \frac{Me_{2n}^{(1)}(u)}{Me_{2n}^{(1)\;\prime}(0)}\simeq&
-\displaystyle\frac{\exp\left(- S(u)\right)}{\left[S'(0) S'(u)\right]^{1/2}} 
-\displaystyle\frac{\ic\pi}{2}\frac{A^{(2n)\; 2}_{0}}{ce_{2n}(\pi/2)^2 } \left[\displaystyle\frac{S'(0)}{ S'(u)}\right]^{1/2}
\cosh\left[S(u)\right]
\label{Langer_Me2n_inside}\\
 \displaystyle\frac{Me_{2n+1}^{(1)}(u)}{Me_{2n+1}^{(1)\;\prime}(0)}\simeq&
-\displaystyle\frac{\exp\left(- S(u)\right)}{\left[S'(0) S'(u)\right]^{1/2}}
-\displaystyle\frac{\ic\pi}{2}\frac{\theta A^{(2n+1)\; 2}_{1}}{ce'_{2n+1}(\pi/2)^2 } \left[\displaystyle\frac{S'(0)}{ S'(u)}\right]^{1/2}
\cosh\left[S(u)\right]\label{Langer_Me2n+1_inside}\ ,
\end{eqnarray}
with $Me_{2n}^{(1)\;\prime}(0)$, and $Me_{2n+1}^{(1)\;\prime}(0)$ being deduced from Eqs.~(\ref{C2n}-\ref{C2n+1}). For the sake of brevity the detailed derivation of Eqs.~(\ref{Langer_Me2n_inside}-\ref{Langer_Me2n+1_inside}) has been put into \ref{WKB_Mathieu_details}.

These formulae are valid when $\frac{2\theta\cosh(2u)}{a_n}<\epsilon_1$, i.e. $2\theta\cosh(2u)\ll a_n$. 
Beyond this condition we locate precisely the turning point by solving
\begin{equation}
  a_n=2\theta\cosh(2u_*)\ .
\end{equation}
Then, in complete similarity with (\ref{Sclass3}), we define the following action:
\begin{equation}
  S(u)-S_*=\displaystyle\int_{u_*}^u\sqrt{|a_{n}-2\theta\cosh(2t)|}\ud t,\quad S_*=S(u_*)\ .
\end{equation}

The WKB approximation as proposed in (\ref{WKB_ex_turn}) gives for the radial Mathieu functions:
\begin{eqnarray}
    \frac{Me^{(1)}_{n}(u)}{Me^{(1)\;\prime}_{n}(0)}\simeq -\sqrt{\pi} \left(\frac{3}{2}\right)^{1/6}e^{-S_*}\frac{\left| S(u)-S_*\right|^{1/6}}{\left[S'(0) |S'(u)|\right]^{1/2}} \times
  \nonumber\\
\left\{\ic \Ai\left[\sgn(u_*-u)\left|\frac{3}{2}\left[S(u)-S_*\right]\right|^{2/3}\right]+
\Bi\left[\sgn(u_*-u)\left|\frac{3}{2}\left[S(u)-S_*\right]\right|^{2/3} \right]\right\}
\label{Langer_Men_cross}
\end{eqnarray}
A detailed derivation of (\ref{Langer_Men_cross}) can be found in \ref{WKB_Mathieu_details}.
When evaluated numerically, the prefactor in (\ref{Langer_Men_cross}) shows a singularity at $u=u_*$. This comes from the fact that both the numerator and the denominator vanish at this point. We removed this artificial divergence using the smooth leading term of the Taylor expansion.
For our purpose, especially to apply it for scattering problems, the approximation (\ref{Langer_Men_cross}) was accurate enough to investigate the far field. Hence we did not have to implement (\ref{WKB_outside}) to approximate radial Mathieu functions.
Last, once 
Eqs~(\ref{Langer_Me2n_inside},\ref{Langer_Me2n+1_inside},\ref{Langer_Men_cross}) are known, $Ce_n(u)$ can be deduced taking the imaginary part.

\subsubsection{WKB formulae for the odd Mathieu functions}%{Other symmetry class}

The same method in order to derive WKB approximation can be followed for the odd symmetry class. 
The classical action to be considered is now:
\begin{equation}
  \label{defS_odd}
  S(u)=\int_0^u\sqrt{b_{n}-2\theta\cosh(2t)}\ud t\ .
\end{equation}

One gets then inside the potential barrier (for $2\theta\cosh(2u)\ll b_n$):
\begin{eqnarray}
  \frac{Ne_{2n}^{(1)}(u)}{Ne_{2n}^{(1)}(0)}&\simeq
\left[\displaystyle\frac{S'(0)}{S'(u)}\right]^{1/2} \exp(-S(u))
 + \displaystyle\frac{\ic\pi \theta^2 B^{(2n)\; 2}_{2}  }{2 se'_{2n}(\pi/2)^2}
 \displaystyle\frac{\sinh(S(u))}{ \left[S'(0)S'(u)\right]^{1/2}}
\label{Langer_Ne2n_inside}\\
 \frac{Ne_{2n+1}^{(1)}(u)}{Ne_{2n+1}^{(1)}(0)}&\simeq
\left[\displaystyle\frac{S'(0)}{S'(u)}\right]^{1/2} \exp(-S(u))
+\displaystyle\frac{\ic\pi \theta B^{(2n+1)\; 2}_{1} }{2 se_{2n+1}(\pi/2)^2}
 \displaystyle\frac{\sinh(S(u))}{ \left[S'(0)S'(u)\right]^{1/2}}
\label{Langer_Ne2n+1_inside}
\end{eqnarray}
with $Ne_{2n}^{(1)}(0)$, and $Ne_{2n+1}^{(1)}(0)$ being deduced from Eqs.~(\ref{D2n}-\ref{D2n+1}).
Similarly as above the details how to derive 
The formulae around the turning point are, for $2\theta\cosh(2u)\approx b_n$:
\begin{eqnarray}
\frac{Ne^{(1)}_{n}(u)}{Ne^{(1)}_{n}(0)}\simeq \sqrt{\pi} \left(\frac{3}{2}\right)^{1/6}e^{-S_*} |S(u)-S_*|^{1/6}\left[\frac{S'(0)}{S'(u)}\right]^{1/2}\times\nonumber\\
\left\{\ic \Ai\left[\sgn(u_*-u)\left|\frac{3}{2}\left[S(u)-S_*\right]\right|^{2/3}\right]+
\Bi\left[\sgn(u_*-u)\left|\frac{3}{2}\left[S(u)-S_*\right]\right|^{2/3} \right]\right\}
\label{Langer_Nen_cross}
\end{eqnarray}
Similarly as above the details how to derive Eqs~(\ref{Langer_Ne2n_inside},\ref{Langer_Ne2n+1_inside},\ref{Langer_Nen_cross}) have been put into \ref{WKB_Mathieu_details}.
Note that, once 
%Eqs~(\ref{Langer_Ne2n_inside},\ref{Langer_Ne2n+1_inside},\ref{Langer_Nen_cross}) 
those equations
are known, $Se_n(u)$ can be deduced taking the imaginary part.
\subsection{Description of our numerical algorithm}

The algorithm works as follow. Before running the program, 
the $A_p^{(n)}$ and $B_p^{(n)}$ are solved, once and for all, using a tri-diagonal matrix diagonalisation algorithm, see \ref{Fourier_coeff}. The results are stored with double precision. In our case, $p_{max}$ and $n_{max}$ has been taken to $200$ and $100$ respectively. The diagonalisation process also provides the characteristic values of the Mathieu equations, $a_n$ and $b_n$, which are also stored. Note that the diagonalisation algorithm does not necessarily give the proper normalization to the $A_p^{(n)}$ and $B_p^{(n)}$. Therefore, Eqs.~(\ref{normAn}) and (\ref{normBn}) need to be fulfilled. 

Once those values have been stored, one can evaluate the radial Mathieu functions. The formula one has to use depends on two considerations: the precision one wants and the distance between $2\theta\cosh(u)$ and $a_n$ (resp. $b_n$ for the odd symmetry class). Indeed, the WKB approximation is valid up to an error scaling as $a_n^{-1/2}$ (resp. $b_n^{-1/2}$). Therefore, the exact series expansion %(\ref{Me2n_seriesfin,Me2n+1_seriesfin,Ne2n_seriesfin,Ne2n+1_seriesfin}) 
(\ref{Me2n_seriesfin}$-$\ref{Ne2n_seriesfin})
are used for small values of $a_n$ (resp. $b_n$). The distance between $2\theta\cosh(u)$ and $a_n$ (resp. $b_n$) determines which formula of the WKB approximation is relevant. 
Note that one very crucial ingredient for the WKB method is to have a large characteristic value. Therefore we restricted ourselves to use the WKB formulae only for $n\ge n_0$. The parameter $n_0$ was taken to $6$ to get a good numerical accuracy.
The algorithm works as follow for the even radial Mathieu function: 
\begin{itemize}
\item $n<n_0$ 
\textbf{OR} $a_n^{-1} \geq \epsilon_0$: Series expansions are used. This case corresponds to small values of $n$ or/and $u$. In this condition, $\epsilon_0$ is the required precision, set to $0.005$ in our case. %$5\ 10^{-3}$
\item {$n>n_0$} \textbf{AND} $a_n^{-1} < \epsilon_0$: Two cases need to be distinguished:
  \begin{itemize}
  \item $2\theta\cosh(u)/a_n < \epsilon_1$: In this case, the formulae (\ref{Langer_Me2n_inside},\ref{Langer_Me2n+1_inside}) are used.
  \item $2\theta\cosh(u)/a_n \geq \epsilon_1$: In this case, the formula (\ref{Langer_Men_cross}) is used. 	
  \end{itemize}
  Our simulations use $\epsilon_1 = 0.1$.
\end{itemize}
Notice that we did not use the third approximation provided by WKB theory, see (\ref{WKB_outside}) and (\ref{WKB_ex_out}). The reason is that, for this special case of the potential, the exact formulae (\ref{Me2n_seriesfin}), and (\ref{Me2n+1_seriesfin}) are then numerically fast and accurate.
The algorithm is the same for the odd Mathieu functions. Only $a_n$ should be changed with $b_n$.
Regarding the choice for $\epsilon_0$ and $\epsilon_1$, several values have been tested to ensure a suitable precision for the algorithm. We took as a benchmark the case $\theta = \pi^2$ for $n = 20$ and we choose $\epsilon_0 = 0.005$ and $\epsilon_1 = 0.1$. 
Note that a different benchmark may lead to a slight change of both $\epsilon_0$ and $\epsilon_1$.

All the results shown in this article have been obtained using a desktop computer running Ubuntu 14.04. The computer uses an Intel$^{\mbox{\scriptsize{\textregistered}}}$ Core{\scriptsize{\texttrademark}} i7-5500U CPU @ $4\times2.4$GHz with a random access memory of 8Gbit. Put aside the computation of the $A_p^{(n)}$ and $B_p^{(n)}$ coefficients, which is made with Mathematica, each formula has been implemented in C/C++. The typical computation time under those circumstances range from a few seconds for Figs~\ref{Me_expu} and \ref{Ne_expu} up to a few minutes for Figs~\ref{GreenN}, \ref{GreenN_vs_a}, \ref{GreenD} and \ref{GreenD_vs_a}.
\subsection{Checking the accuracy of the WKB formulae}

In this Section we first display the results obtained via the method exposed previously. Secondly the WKB formulae are compared with two other estimates. 
The first estimate is only valid for the radial Mathieu functions of the first kind and is computed via Mathematica by using Eq.~(\ref{Ang_to_rad}).
The second estimate is provided by a standard numerical library \cite{Zhang}.

\begin{figure}[!ht]
  \includegraphics[width=\linewidth]{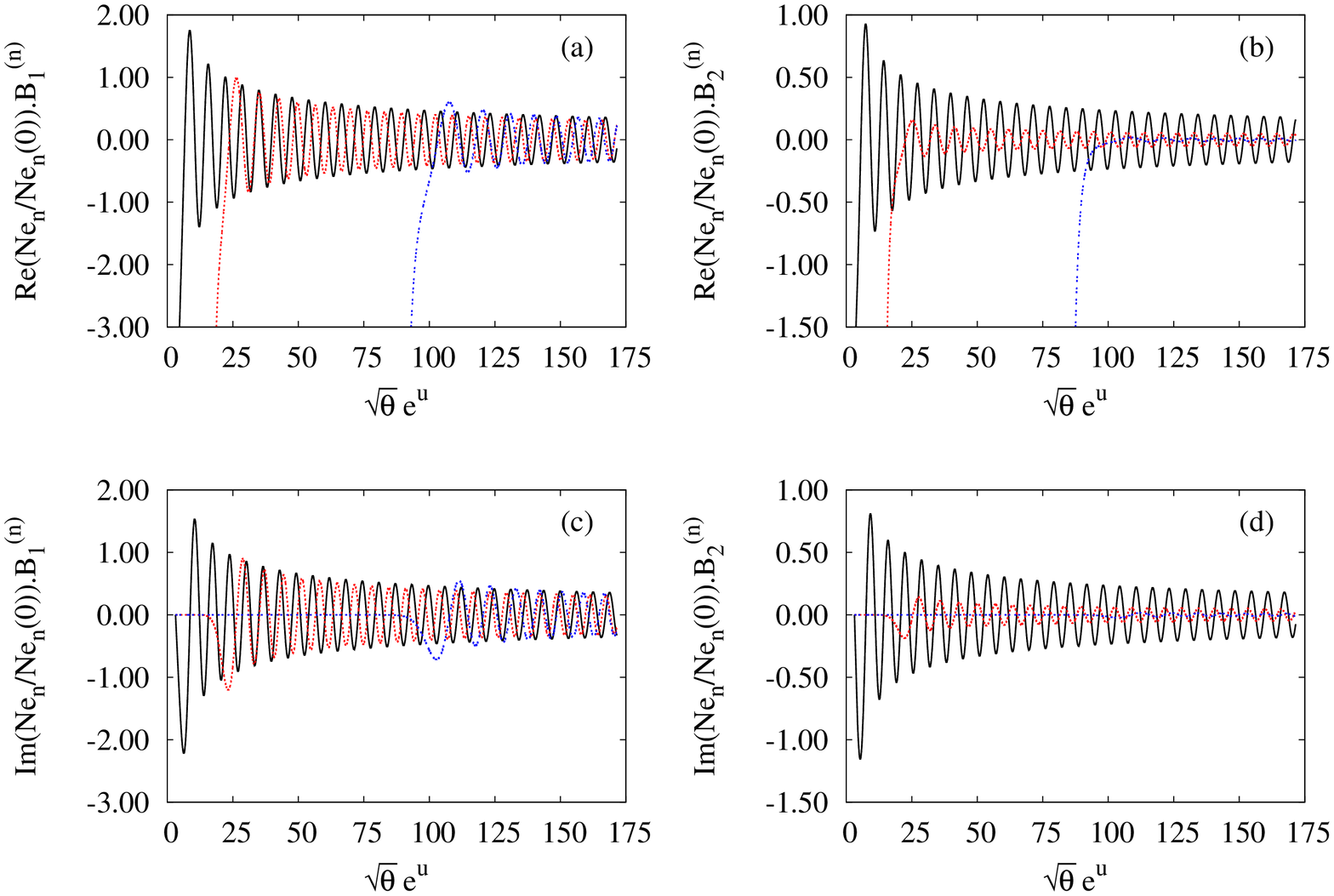}
  \caption{Radial Mathieu functions of the third kind for $a/\lambda=2$, and for the odd symmetry class as a function of the ratio between the polar radius and the wavelength, see (\ref{radius_ell}). Top: real part. Bottom: imaginary part. (a) and (c): black: $n=5$, red: $n=21$, blue: $n=99$.
(b) and (d): black: $n=4$, red: $n=20$, blue: $n=100$.}
  \label{Ne_expu}
\end{figure}

A more intuitive way to represent radial Mathieu functions is to plot them as a function of the ratio between the polar radius and the wave length using:
\begin{equation}
  k|{\bf x}| \sim \frac{ka}{4} e^u=\sqrt{\theta}e^u\label{radius_ell}\ .
\end{equation}
This is particularly useful to estimate whether one can also explore the far field with our method (see below). Such plots are displayed in Figs.~\ref{Me_expu} and \ref{Ne_expu}. They show first that our method works equally for both symmetry classes. Secondly it is also adequate to explore the far field regime.
\begin{figure}[!ht]
  \includegraphics[width=\linewidth]{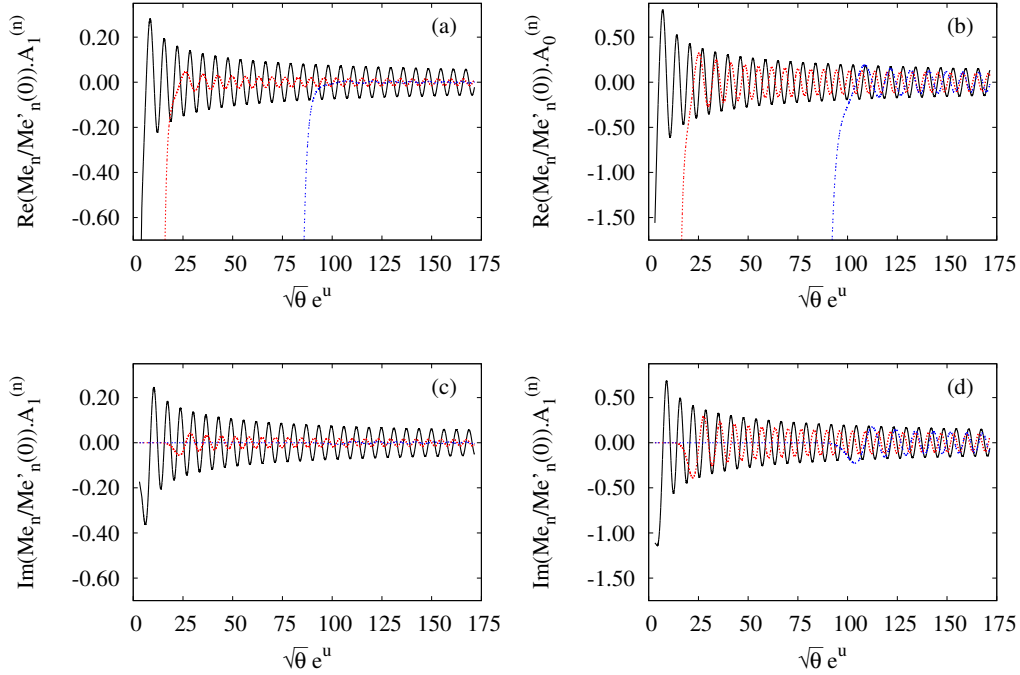}
  \caption{Radial Mathieu functions of the third kind for $a/\lambda=2$, and for the even symmetry class as a function of the ratio between the polar radius and the wavelength, see (\ref{radius_ell}). Top: real part. Bottom: imaginary part. (a) and (c): black: $n=5$, red: $n=21$, blue: $n=99$.
(b) and (d): black: $n=4$, red: $n=20$, blue: $n=100$.}
  \label{Me_expu}
\end{figure}

\begin{figure}[!ht]
  \includegraphics[width=\linewidth]{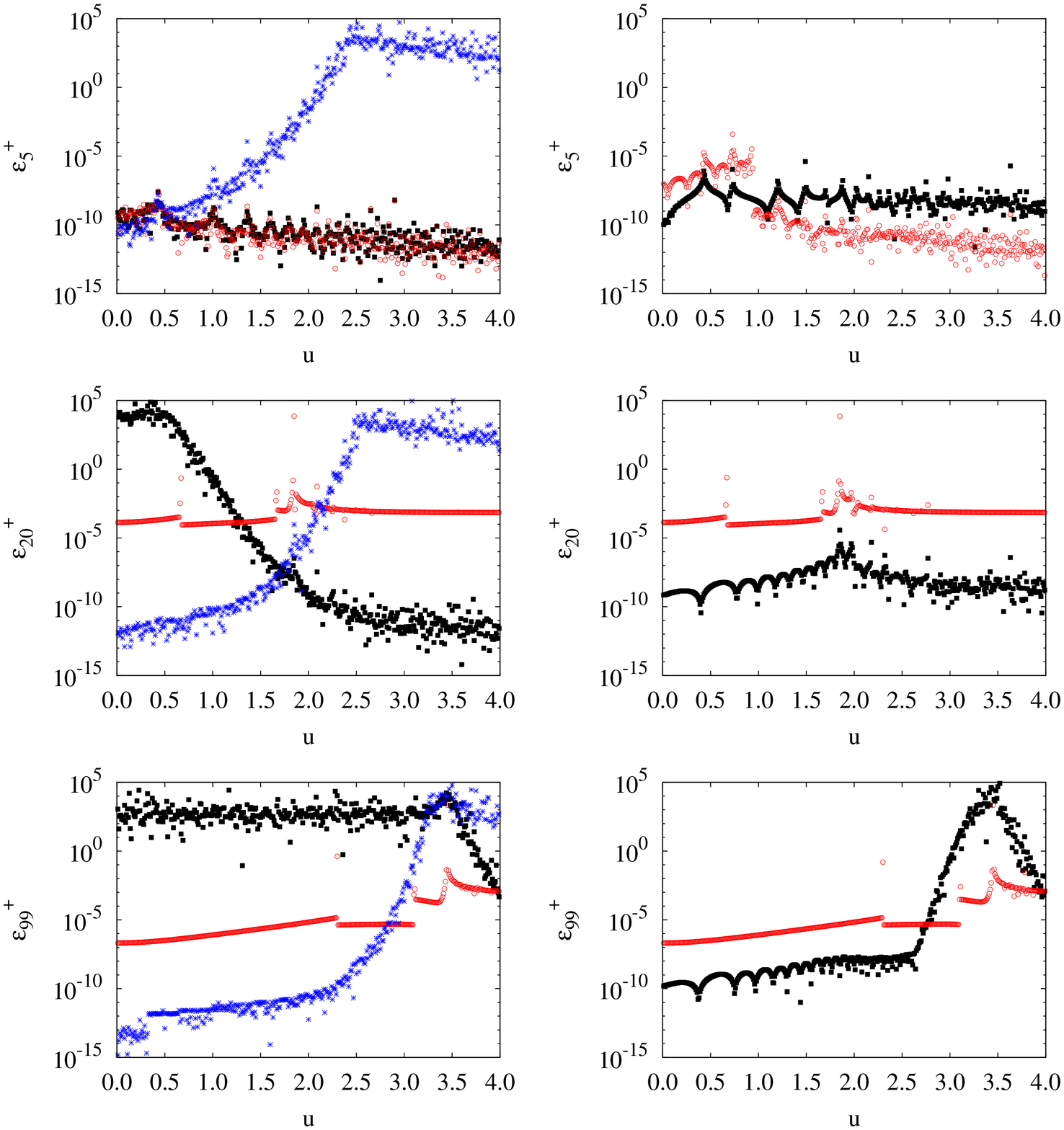}
\caption{
Relative error for the estimate of radial Mathieu function for $a/\lambda=2$ for the even symmetry class, of the first kind (left column), and of second kind (right column). Red circles stand for the results obtained via our WKB-like method. Black squares stand for the estimate of a standard numerical library. Blue stars comes from Eq.~(\ref{Ang_to_rad}) using Mathematica. 
}
  \label{series_vs_WKB_e}
\end{figure}

\begin{figure}[!ht]
  \includegraphics[width=\linewidth]{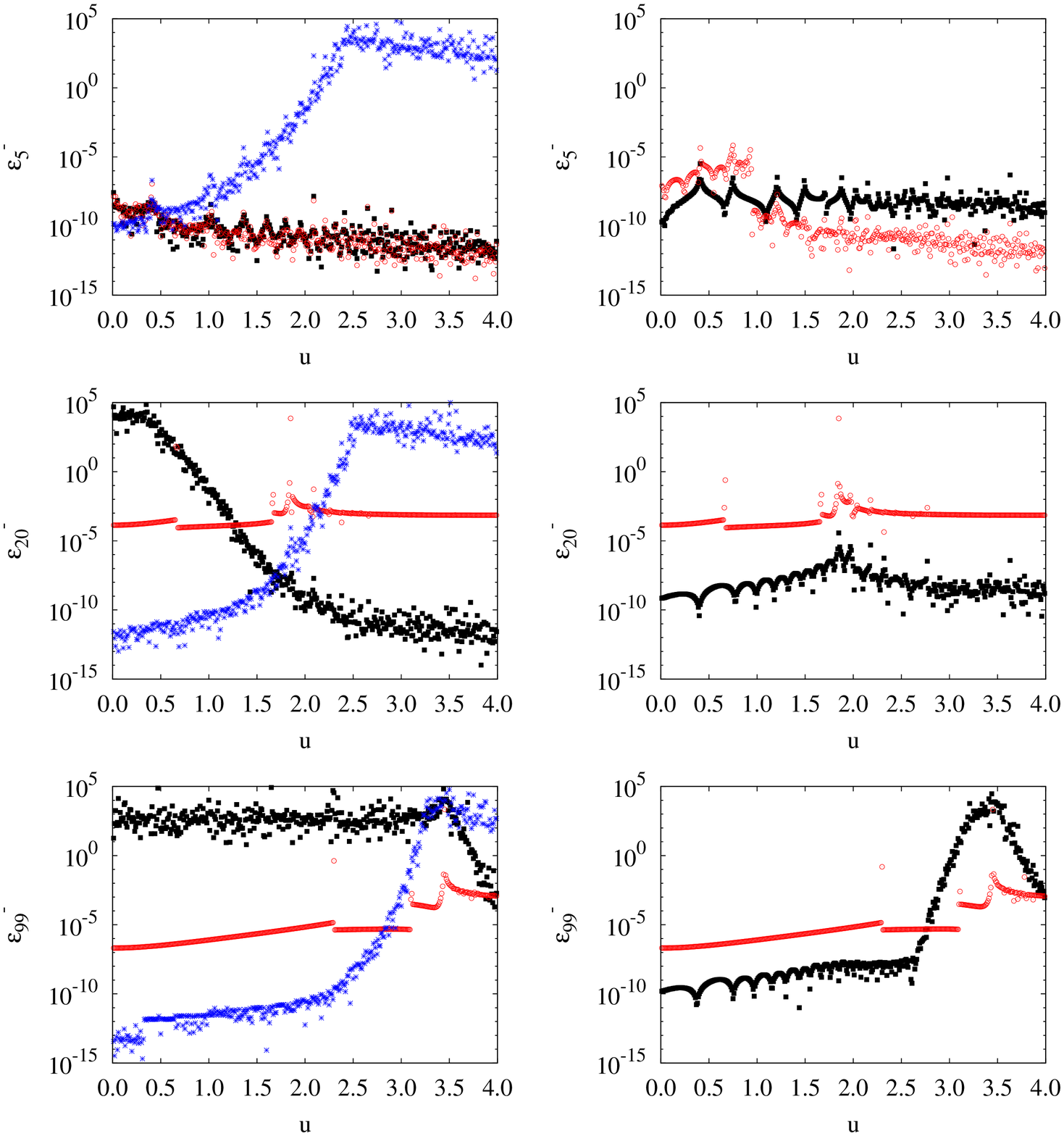}
\caption{
Relative error for the estimate of radial Mathieu function for $a/\lambda=2$ for the odd symmetry class, of the first kind (left column), and of second kind (right column). Red circles stand for the results obtained via our WKB-like method. Black squares stand for the estimate of a standard numerical library. Blue stars comes from Eq.~(\ref{Ang_to_rad}) using Mathematica. 
}
  \label{series_vs_WKB_o}
\end{figure}

%First it is illustrated how the WKB approach is compared to the expansion into product of Bessel functions in
Then our results are checked by considering the following quantity:
\begin{equation}
  \epsilon^{+}_n\equiv \left|\frac{\frac{Ce_n''(u)}{Ce_n(u)}  - \left[a_n -2\theta\cosh(2u)\right]}{a_n -2\theta\cosh(2u)}\right|,\quad
  \epsilon^{-}_n\equiv \left|\frac{\frac{Se_n''(u)}{Se_n(u)}  - \left[b_n -2\theta\cosh(2u)\right]}{b_n -2\theta\cosh(2u)}\right|,
\label{epsn}
\end{equation}
which are computed in three ways in Fig.~\ref{series_vs_WKB_e}, and \ref{series_vs_WKB_o}. Those quantities are especially relevant as they are not sensitive to the choice of overall normalization used for the Mathieu function. Note that they were used to check third kind radial Mathieu functions by replacing $Ce_n(u)$ with $\textrm{Im }Me_n^{(1)}(u)$, and $Se_n(u)$ with $\textrm{Im }Ne_n^{(1)}(u)$.
The quantities (\ref{epsn}) can be viewed as the relative error estimate on the computation of the radial Mathieu functions.

The first more naive observation is that Eq.~(\ref{Ang_to_rad}) is not an adequate way to estimate the radial Mathieu functions. The second observation is that our method is more accurate than the numerical library \cite{Zhang} for small values of $u$ (corresponding to near field in the scattering problem). This claim is better illustrated when the index $n$ increases. We also notice that the use of smooth explicit functions gives in general smoother variations of the errors as compared to the numerical library. It is also worth pointing that there is a clear local increase of the relative error for our results exactly at the region $u=u_*$. Again this is in line with the fact that WKB method is known to work less accurately near a turning point. Another benefit of our approach is that it is especially suited and very efficient for series of Mathieu functions, see below and \ref{check_identity}.
  As a conclusion, the general relative error of our results can be estimated around $10^{-4}$ and is better to a standard numerical library in several instances. Crucially we only considered here the leading term of the WKB expansion. Therefore we expect the accuracy of our method to be significantly increased when including higher order terms in the series.
\section{Application: Green function for scattering problems.}
\label{Green}

In this Section, we applied our method to evaluate Mathieu functions to a scattering problem. The goal is to compute numerically the Green function for a wave scattered by a single slit. First the formula for the Green function as a series of Mathieu functions is recalled. Then we show numerical evaluation of the Green function following the WKB method exposed in the previous Section. 
The results are shown successively for Neumann boundary condition, i.e. when the wave amplitude is constrained to have a vanishing normal derivative along the arms of the slit, and for Dirichlet boundary condition, i.e. when the wave amplitude vanishes on both arms. 
In \ref{check_identity} it is checked via a mathematical identity that our method is also suitable for series of Mathieu functions.

\subsection{Green function for the scattering by a slit with Neumann boundary condition}

The series expansion of the Green function for this problem has been derived (using other conventions for Mathieu functions) more than a century ago \cite{Sieger}. Therefore the full derivation of this formula with a modern notation is moved to \ref{derive_Green}. Assuming $v_0<0$ the final expression is:
\begin{equation}
  \label{GsingleslitN_v2}
\hspace{-2cm}G({\bf x},{\bf x}_0,k)=\left\{
\begin{array}{lc}
\displaystyle\frac{1}{\pi}\displaystyle\sum_{n\ge 0} \frac{Me^{(1)}_n(u)}{Me^{(1)\;\prime}_n(0)} \frac{Me^{(1)}_n(u_0)}{Me^{(1)}_n(0)}ce_n(v)ce_n(v_0)& 0<v< \pi\\ & \\
\displaystyle\frac{1}{\pi}\displaystyle\sum_{n\ge 0} \left[2\frac{Me^{(1)}_n(u_>)}{Me^{(1)\;\prime}_n(0) }\frac{Ce_n(u_<)}{ce_n(0) } -
\frac{Me^{(1)}_n(u)}{Me^{(1)\;\prime}_n(0)} \frac{Me^{(1)}_n(u_0)}{Me^{(1)}_n(0)} \right]ce_n(v_0) ce_n(v)& -\pi < v <0
\end{array}\right.\ ,
\end{equation}
where the following notation was introduced:
\begin{eqnarray}
  u_<=\min(u,u_0),\quad u_>=\max(u,u_0)\ .\nonumber
\end{eqnarray}

\begin{figure}[!ht]
  \begin{center}
    \includegraphics[width=\textwidth]{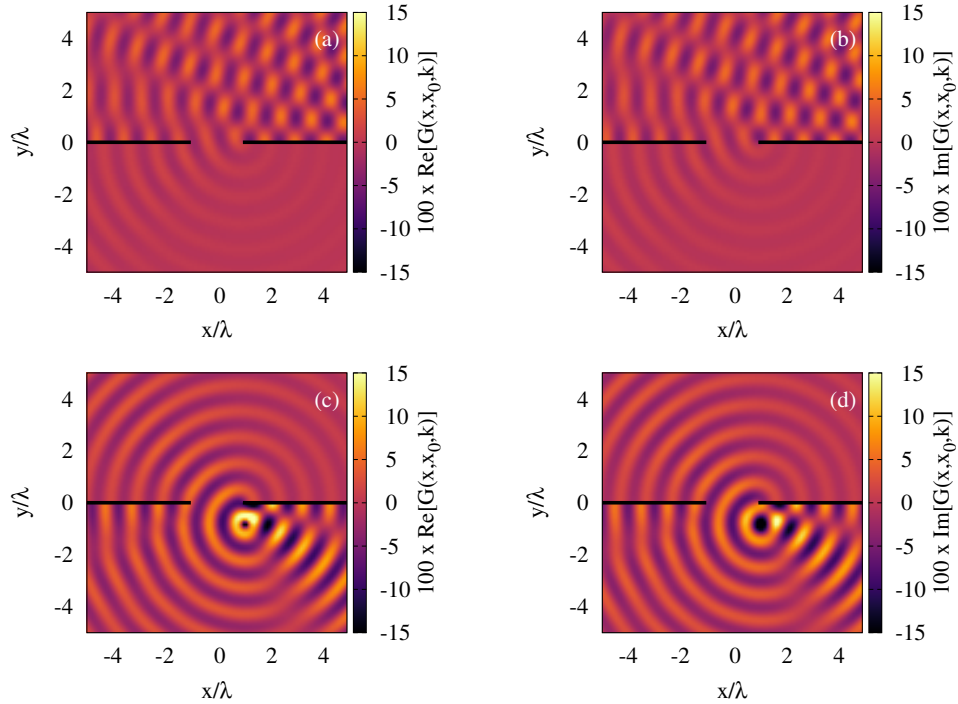}
  \end{center}  
  \caption{Examples of Green function for the scattering by a single slit with Neumann boundary condition and $a/\lambda = 2$. (a) and (b): diffraction for a source at $(x_0,y_0)\simeq(9.301,3.834)$, real and imaginary part respectively. (c) and (d): diffraction for a source at $(x_0,y_0)\simeq(1.091,-0.831)$, real and imaginary part respectively.}
  \label{GreenN}
\end{figure}

\begin{figure}[!ht]
  \begin{center}
    \includegraphics[width=\textwidth]{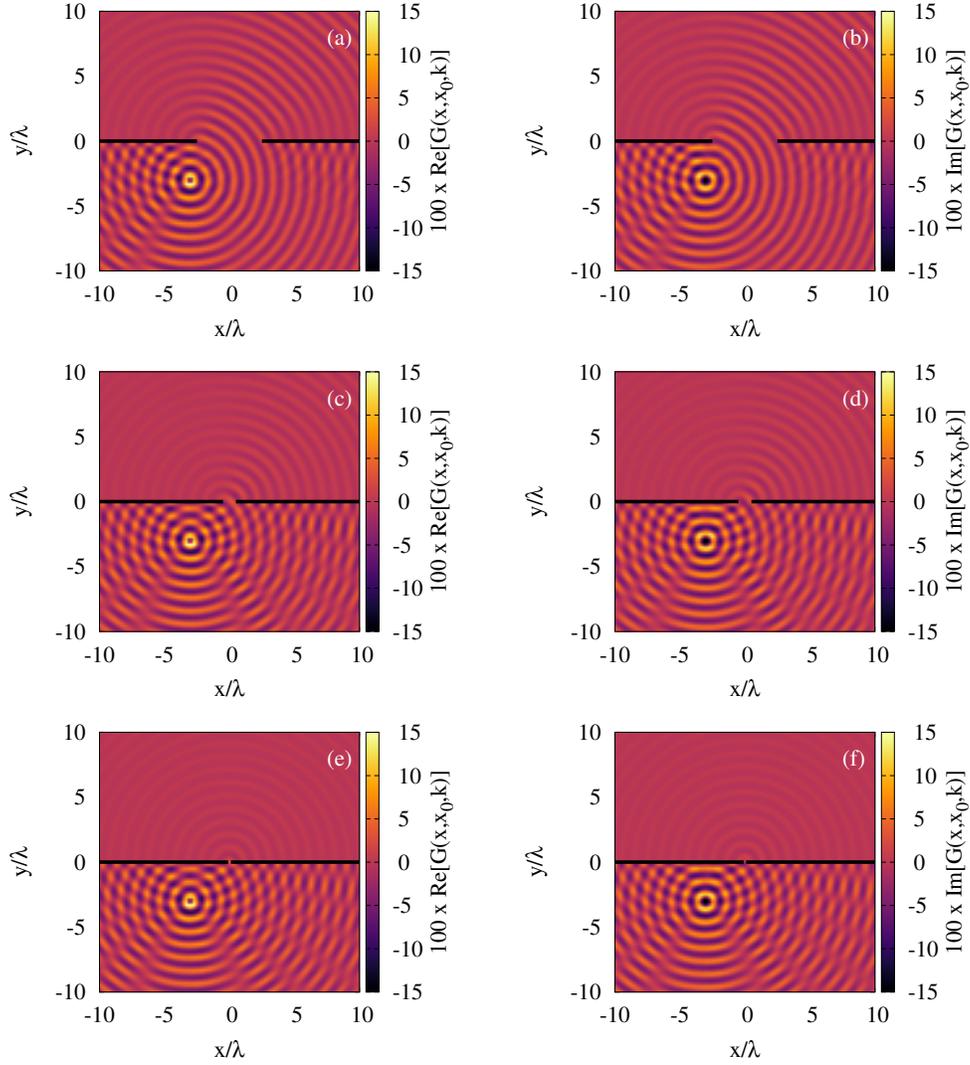}
  \end{center}  
  \caption{Green function for the diffraction, with Neumann boundary condition, by a slit for several slit width: (a) and (b): $a/\lambda = 5$ ($\theta = 25\pi^2/4$) real and imaginary part respectively. (c) and (d): $a/\lambda = 1$ ($\theta = \pi^2/4$) real and imaginary part respectively, (e) and (f): $a/\lambda = 0.2$ ($\theta = \pi^2/100$) real and imaginary part respectively. The source is located at $(x_0,y_0)=(-3,-3)$ in every cases.}
  \label{GreenN_vs_a}
\end{figure}

Evaluating the series of Mathieu function following the previous Sections, the Green function can be estimated, see Fig.~\ref{GreenN}. In particular we stress that our method is able to cover the case of close or distant location of the source for a moderate value of the ratio between the slit width and the wavelength. % It is also useful for both cases when the source location is close or far from the slit.
In Fig.~\ref{GreenN_vs_a} it is shown moreover that our method gives a smooth result for the Green function in a larger range of this ratio.

\subsection{Green function for the scattering by a slit with Dirichlet boundary condition}

Here the method previously described is used to treat the case of another boundary condition, which is more common in quantum problems. Assume $v_0<0$. The derivation of the Green function follows exactly the same steps as described in \ref{derive_Green} so we only give here the result:
\begin{equation}
  \label{GsingleslitD_v2}
\hspace{-2cm}G({\bf x},{\bf x}_0,k)=\left\{
\begin{array}{lc}
-\displaystyle\frac{1}{\pi}\displaystyle\sum_{n> 0} \frac{Ne^{(1)}_n(u)}{Ne^{(1)}_n(0)} \frac{Ne^{(1)}_n(u_0)}{Ne^{(1)\;\prime}_n(0)}se_n(v)se_n(v_0)& 0<v< \pi\\ & \\
\displaystyle\frac{1}{\pi}\displaystyle\sum_{n> 0} \left[-2\frac{Ne^{(1)}_n(u_>)}{Ne^{(1)}_n(0) }\frac{Se_n(u_<)}{Se'_n(0) } +
 \frac{Ne^{(1)}_n(u)}{Ne^{(1)}_n(0)} \frac{Ne^{(1)}_n(u_0)}{Ne^{(1)\;\prime}_n(0)}\right]se_n(v)se_n(v_0)& -\pi < v <0
\end{array}\right.\ .
\end{equation}

\begin{figure}[!ht]
  \begin{center}
    \includegraphics[width=\textwidth]{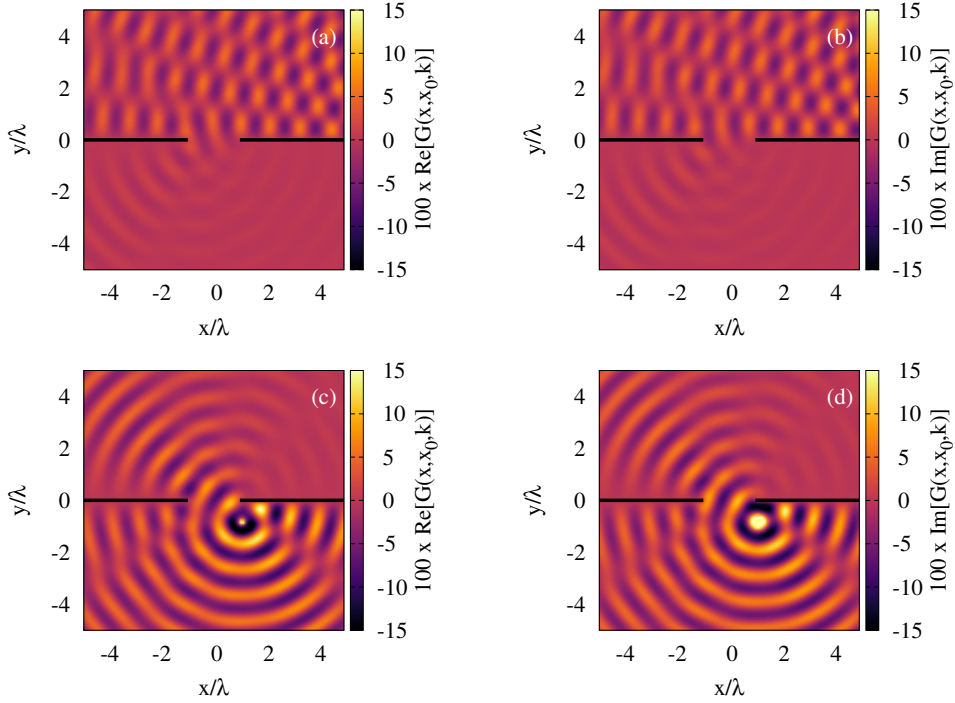}
  \end{center}  
  \caption{Examples of Green function for the scattering by a single slit with Dirichlet boundary condition and $a/\lambda = 2$. (a) and (b): diffraction for a source at $(x_0,y_0)\simeq(9.301,3.834)$, real and imaginary part respectively. (c) and (d): diffraction for a source at $(x_0,y_0)\simeq(1.091,-0.831)$, real and imaginary part respectively.}
  \label{GreenD}
\end{figure}

The results of our computation are displayed in Fig.~\ref{GreenD}, again when the source location is either close or far from the slit.

\begin{figure}[!ht]
  \begin{center}
    \includegraphics[width=\textwidth]{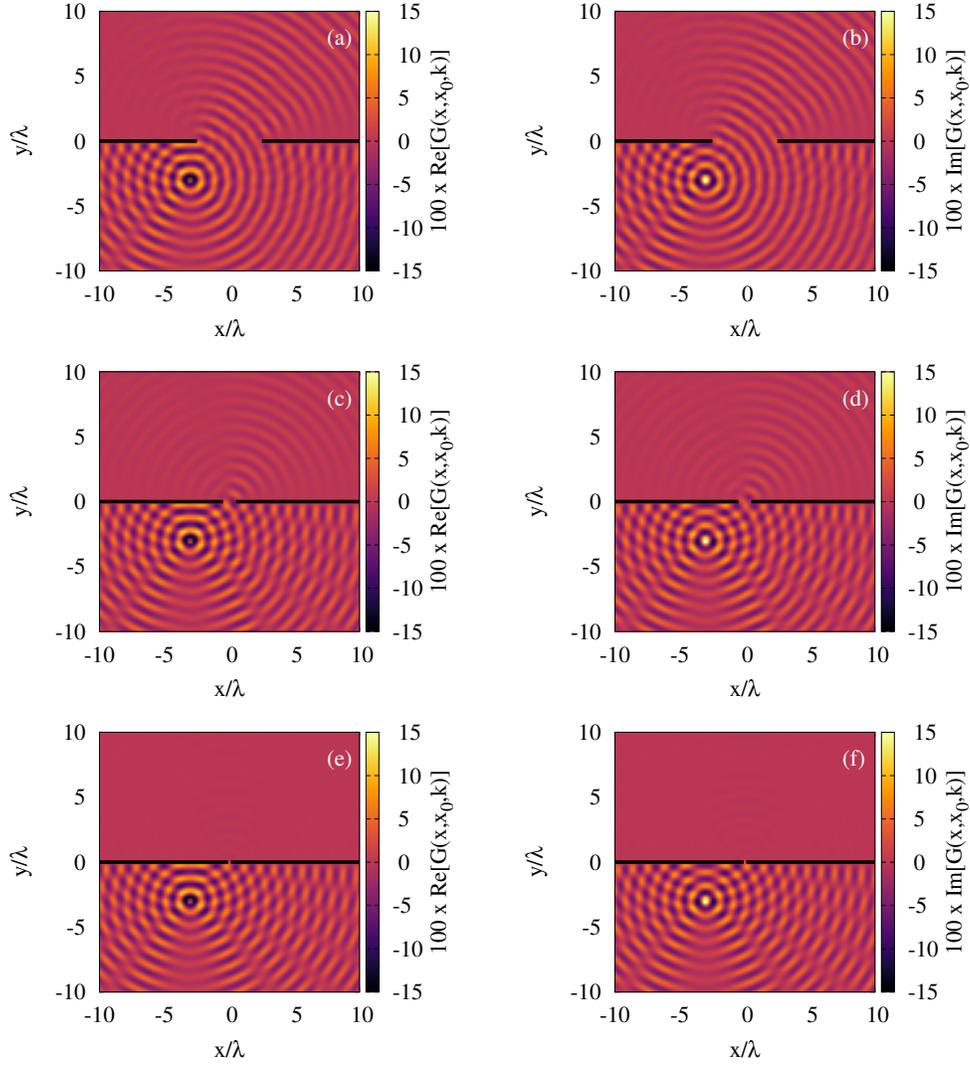}
  \end{center}  
  \caption{Green function for the diffraction, with Dirichlet boundary condition, by a slit for several slit width: (a) and (b): $a/\lambda = 5$ ($\theta = 25\pi^2/4$) real and imaginary part respectively. (c) and (d): $a/\lambda = 1$ ($\theta = \pi^2/4$) real and imaginary part respectively, (e) and (f): $a/\lambda = 0.2$ ($\theta = \pi^2/100$) real and imaginary part respectively. The source is located at $(x_0,y_0)=(-3,-3)$ in every cases.}
  \label{GreenD_vs_a}
\end{figure}

It is also checked in Fig.~\ref{GreenD_vs_a} that our method works equally well when one varies the ratio between the slit width and the wave length.

Another check that we run is to consider the scattering problem by a strip of finite length. Although it can be treated via Babinet's principle when the scattering by a slit has been solved, it is also possible to write down the Green function as a series of Mathieu functions. These expansions are given in \ref{Strip} and it was checked that we also obtain equally reliable pictures of the Green function.

\subsection{Far field. Comparison with Fraunhofer theory}

It is worth comparing our results and check how efficient they can be in the far field, which is a regime easier to access in experiments. 
The far field is defined as the angular distribution of the $f(\alpha)$ for a plane wave ($u_0\gg 1$ in our numerics) incident on the obstacle. It translates in our notation as ($u_m$ is fixed to a large value):
\begin{equation}
  f(\alpha)=G(u_m,v=\alpha,u_0,v_0,k)
\end{equation}

It can be compared to Fraunhofer's formula for far field diffraction:
\begin{equation}
  f_F(\alpha)\propto\frac{\sin\left(2\sqrt{\theta} \sin(\alpha-v_0)\right)}{2\sqrt{\theta}\sin(\alpha-v_0) }
\end{equation}

\begin{figure}[!ht]
  \begin{center}
    \includegraphics[width=\textwidth]{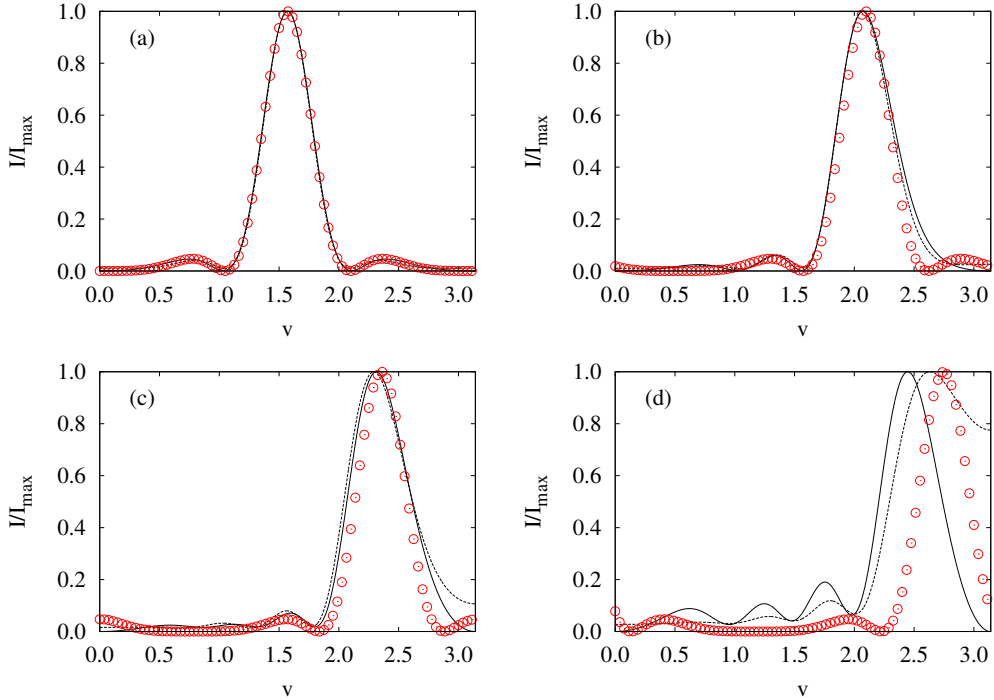}
  \end{center}  
  \caption{Intensity $I(v)  = |G({\bf x},{\bf x}_0,k)|^2$ of the diffraction pattern for the far-field approximation as a function of the angular elliptical coordinate $v$ of ${\bf x}$. The intensity has been scaled by its maximal value $I_{\textrm{max}}$. For each figure $u_0 = u = 5$. (a) $v_0 = \pi/2$, (b) $v_0 = \pi/3$, (c) $v_0 = \pi/4$, (d) $v_0 = \pi/8$. Red dot: Fraunhofer approximation, black solid line: Dirichlet boundary condition, black dashed line: Neumann boundary condition.}
  \label{farfield}
\end{figure}

The far field from our method is displayed in Fig.~\ref{farfield}.
Interestingly we found that not only our computations reproduce Fraunhofer's prediction for an angle of incidence equal to $90^{\circ}$, but they can also span a small angle of incidence (grazing incidence) where clear deviations from Fraunhofer's theory are expected, because the latter prediction does not fulfill the boundary condition at the obstacle as seen in Fig.~\ref{farfield}.
\section{Conclusion}
\label{conclusion}

The study of the scattering of a scalar wave by a single slit, or by a strip, has been considered using the exact solution expressed as series of Mathieu functions. While this approach seems more promising when the wavelength is of the same order as the slit/strip width, the standard way to evaluate Mathieu functions as series of products of Bessel functions is much less accurate. We designed a simple, accurate and fast way to approximate radial Mathieu functions from a WKB perspective. Furthermore our numerical method is steadily implemented on a desktop computer.
The approximating formulae are explicit so they might be also useful for analytical computations. Our formula were compared with existing numerical library and it was shown that our method is more accurate for small argument and moderate or large index.
It was checked that these formulae are also relevant and accurate to estimate series of Mathieu functions in order to obtain the solution of the scattering problem. It is also worth adding that the method here applies also for the solution of radial Mathieu equation with arbitrary (large) values of $h$. At last a very stimulating extension of our method consists of using recent advances to extend WKB solutions in the complex plane \cite{Shudo} in order to compute resonances of the scattering problem.

While we restrict ourselves to the leading order to WKB expansions a stimulating perspective is to consider systematic increase of the current method by accounting for higher order terms. This stands beyond the scope of this paper.
Another direct use of our method is to consider more complex geometries, e.g. when the wave is scattered by an obstacle, whose boundary is only piecewise a conic.
Finally we hope that the explicit formulae we used for radial Mathieu functions may be also useful for other physical situations, e.g. in quantum defect theory.

\section*{Acknowledgment}

The authors would like to thank T. Bastin, J. Martin, N. Vandewalle and P. Schlagheck for encouraging us with their stimulating remarks throughout the redaction. R.D. also wants to thank J.~Shim for his careful reading of the manuscript and H. Landa for providing Refs. \cite{OMalley,Hunter,Gao}.
\pagebreak
%% The Appendices part is started with the command \appendix;
%% appendix sections are then done as normal sections
\appendix

\section{Change of normalization/notation}
\label{notations}
%We first want to briefly give the correspondence between our notation and some others commonly seen in the litterature:
Here the correspondence is given between our notation and some others sometimes encountered in the literature:
\bigskip

\begin{tabular}[c]{|p{3cm}|p{4cm}|p{5cm}|}
  \hline
  Present notation & Notation in %Abramovitz and Stegun 
\cite{AbramowitzStegun} & Notation in %Gradshteyn and  Ryzhik 
\cite{GradRiz}\\
  \hline
  $\theta$ & $q$ & $q$\\
\hline
$a_n(\theta)$ & $a_n(q)$ & $a_n(q)$\\
\hline
$b_n(\theta)$ & $b_n(q)$ & $b_n(q)$\\
\hline
$Ce_n(\theta,z)$ & $Ce_n(z,q)$ & $Ce_n(z,q)$\\
\hline
$Se_n(\theta,z)$ & $Se_n(z,q)$ & $Se_n(z,q)$\\
\hline
$Me^{(1)}_{2n}(\theta,z)$ & $\alpha_{2n} Mc^{(3)}_{2n}(z,q)$ & $Ce_{2n}(z,q)+\ic Fey_{2n}(z,q)$\\
\hline
$Me^{(1)}_{2n+1}(\theta,z)$ & $\alpha_{2n+1} Mc^{(3)}_{2n+1}(z,q)$ & $Ce_{2n+1}(z,q)+\ic Fey_{2n+1}(z,q)$\\
\hline
$Me^{(2)}_{2n}(\theta,z)$ & $\alpha_{2n} Mc^{(4)}_{2n}(z,q)$ & $Ce_{2n}(z,q)-\ic Fey_{2n}(z,q)$\\
\hline
$Me^{(2)}_{2n+1}(\theta,z)$ & $\alpha_{2n+1} Mc^{(4)}_{2n+1}(z,q)$ & $Ce_{2n+1}(z,q)-\ic Fey_{2n+1}(z,q)$\\
\hline
$Ne^{(1)}_{2n}(\theta,z)$ & $\beta_{2n} Ms^{(3)}_{2n}(z,q)$ & $Se_{2n}(z,q)+\ic Gey_{2n}(z,q)$\\
\hline
$Ne^{(1)}_{2n+1}(\theta,z)$ & $\beta_{2n+1} Ms^{(3)}_{2n+1}(z,q)$ & $Se_{2n+1}(z,q)+\ic Gey_{2n+1}(z,q)$\\
\hline
$Ne^{(2)}_{2n}(\theta,z)$ & $\beta_{2n} Ms^{(4)}_{2n}(z,q)$ & $Se_{2n}(z,q)-\ic Gey_{2n}(z,q)$\\
\hline
$Ne^{(2)}_{2n+1}(\theta,z)$ & $\beta_{2n+1} Ms^{(4)}_{2n+1}(z,q)$ & $Se_{2n+1}(z,q)-\ic Gey_{2n+1}(z,q)$\\
\hline
% \caption{Correspondence between different notations for the Mathieu functions}
\end{tabular}

with the following factors
\begin{eqnarray*}
  \alpha_{2n}&=&\frac{(-1)^n ce_{2n}(\theta,\pi/2) ce_{2n}(\theta,0)}{ A^{(2n)}_{0}(\theta)}\\
  \alpha_{2n+1}&=&-\frac{ (-1)^n ce'_{2n+1}(\theta,\pi/2) ce_{2n+1}(\theta,0) }{\sqrt{\theta} A^{(2n+1)}_{1}(\theta)}\\
  \beta_{2n}&=& -\frac{(-1)^nse_{2n+1}(\theta,\pi/2) se'_{2n+1}(\theta,0)}{ \sqrt{\theta}B^{(2n+1)}_{1}(\theta)}\\
  \beta_{2n+1}&=&\frac{(-1)^n se'_{2n}(\theta,\pi/2) se'_{2n}(\theta,0)}{\theta B^{(2n)}_{2}(\theta)}
\end{eqnarray*}

% \section{Connecting the WKB approximations for the $\cosh$ potential}
% \label{connectionWKB}

% To be written
\section{Fourier coefficients for the angular Mathieu functions}
\label{Fourier_coeff}

Here is detailed how to obtain the Fourier coefficients for angular Mathieu functions. The method is quite standard so it is first described for one given symmetry class. The required recursion relations for the other symmetry classes are then simply recalled.

When putting the expansion (\ref{Fourier_ce_even}) into (\ref{eqmathieu}) one gets the following recursion relations:
\begin{eqnarray}
  a_{2n}  A^{(2n)}_{0} - \theta A^{(2n)}_{2} &=&0\nonumber\\
  \left[ a_{2n}-4\right]A^{(2n)}_{2} -2\theta A^{(2n)}_{0} -A^{(2n)}_{4}(\theta) &=&0\nonumber\\
  \forall p\ge 2,\quad \left[ a_{2n}-4p^2\right]A^{(2n)}_{2p} -\theta\left[A^{(2n)}_{2p-2}+A^{(2n)}_{2p+2} \right]&=&0
  \label{recurs_A2n}
\end{eqnarray}
This means that the coefficients $A^{(2n)}_{2p}$ can be seen as the entries of the eigenvector of a (infinite dimensional) tridiagonal matrix associated to the eigenvalues $a_{2n}$. The eigenvalue/eigenvector problem for such matrices is very efficiently solved by standard software or libraries. Note that one should also fulfill the normalization conditions (\ref{normAn}).
We decided to use Mathematica to solve the eigenvalue problem, and stored the coefficients with normal double precision ($16$ Digits).

The recursion relations for the Fourier coefficients of the even angular Mathieu functions with odd index are obtained when inserting the expansion (\ref{Fourier_ce_odd}) into (\ref{eqmathieu}):
\begin{eqnarray}
  \left[a_{2n+1}-\theta-1\right]  A^{(2n+1)}_{1} - \theta A^{(2n+1)}_{3} &=0\nonumber\\
  \forall p\ge 1, \left[ a_{2n+1}-(2p+1)^2\right]A^{(2n+1)}_{2p+1} -\theta\left[A^{(2n+1)}_{2p-1}+A^{(2n+1)}_{2p+3} \right]&=0
  \label{recurs_A2n+1}
\end{eqnarray}
The recursion relations for the Fourier coefficients of the odd angular Mathieu functions with odd index are obtained when inserting the expansion (\ref{Fourier_se_odd}) into (\ref{eqmathieu}):
\begin{eqnarray}
  \left[b_{2n+1}+\theta-1\right]  B^{(2n+1)}_{1} - \theta B^{(2n+1)}_{3} &=0\nonumber\\
  \forall p\ge 1, \left[ b_{2n+1}-(2p+1)^2\right]B^{(2n+1)}_{2p+1} -\theta\left[B^{(2n+1)}_{2p-1}+B^{(2n+1)}_{2p+3} \right]&=0
  \label{recurs_B2n+1}
\end{eqnarray}
The recursion relations for the Fourier coefficients of the odd angular Mathieu functions with odd index are obtained when inserting the expansion (\ref{Fourier_se_even}) into (\ref{eqmathieu}):
\begin{eqnarray}
  \left[b_{2n}-4\right]  B^{(2n)}_{2}- \theta B^{(2n)}_{4} &=0\nonumber\\
  \forall p\ge 2, \left[ b_{2n}-4p^2\right]B^{(2n)}_{2p} -\theta\left[B^{(2n)}_{2p-2}+B^{(2n)}_{2p+2} \right]&=0
  \label{recurs_B2n}
\end{eqnarray}

\section{Detailed derivation of WKB formulae for radial Mathieu functions}
\label{WKB_Mathieu_details}

In this Section the use of WKB is exposed in full length. This aims to make this method more accessible to the reader and explain how the formulae for the main text have been derived.

The main idea is that in all general WKB formulae (\ref{WKB_inside}), (\ref{WKB_outside}) and (\ref{WKB_turning}) there are two undetermined coefficients. First the continuity of both the wave function and its derivative is always assumed when changing the regime, depending on $E-V(x)$ is positive, negative or has a zero.
One is left in general with only two overall undetermined constants. Those are fixed by boundary condition imposed to the considered solution of Mathieu functions.

An important remark about WKB method is that it consists of writing the approximation of a solution of modified Mathieu equation as an \emph{asymptotic} expansion. In particular this means that, in a more rigorous and systematic approach, see e.g.~\cite{Dingle}, exponentially small terms have to be cancelled in the expansion. Here we will expose how we chose to keep some of them, solely based on an numerical efficiency consideration. 

Below the main WKB formula for third kind Mathieu functions is derived for each symmetry class successively. From that we could obtain expression, which is especially efficient for small $u$. The corresponding formulae for the first kind radial Mathieu functions are easily obtained by taking the real part. 

\subsection{WKB approximations for $Me^{(1)}_n(u)$}

In the case of the odd symmetry class $Me^{(1)}_n(u)$ is defined as the unique complex-valued solution of (\ref{eqmathieum}) which obeys the following requirements:
\begin{itemize}
\item it obeys outgoing radiation condition at infinity,
\item its real part is $Ce_n(u)$.
\end{itemize}
Let us start with the formula (\ref{Langer_Men_cross}). It will be obtained from (\ref{WKB_turning}) by adjusting $C_+$, and $C_-$ in order to fulfill the above-listed requirements. First, the zero of $E-V(x)$ is crossed in such a way that the minus sign has to be chosen. Second we write its large argument asymptotics using the standard formulae for Airy functions:
\begin{equation}
  \label{asympt_Airy-}
  \Ai(-x) \sim  \frac{\cos\left(\frac{2}{3}(-x)^{3/2}-\frac{\pi}{4}\right)}{\sqrt{\pi} (-x)^{1/4}},\
  \Bi(-x) \sim -\frac{\sin\left(\frac{2}{3}(-x)^{3/2}-\frac{\pi}{4}\right)}{\sqrt{\pi} (-x)^{1/4}},\ 
x\to\infty
\end{equation}
Inserting (\ref{asympt_Airy-}) into (\ref{WKB_turning}) leads to when $|S(u)-S_*|=S(u)-S_*\to\infty$:
\begin{equation}
  \frac{Me^{(1)}_n(u)}{Me^{(1)\;\prime}_n(0)}\sim \frac{1}{\sqrt{\pi}\left(\frac{3}{2}\right)^{\frac{1}{6}} [S'(u)]^{\frac{1}{2}}}
\left[ C_+ \cos\left(S(u)-S_*-\frac{\pi}{4}\right)-C_- \sin\left(S(u)-S_*-\frac{\pi}{4}\right)\right]
\end{equation}
For large argument, $S(u)\sim \sqrt{\theta} e^u$ hence the radiation condition at infinity reads:
\begin{equation}
  C_+=\ic C_-\ . \label{C+-}
\end{equation}
Rewriting (\ref{WKB_turning}) with (\ref{C+-}) leaves one constant $C$ to be determined:
\begin{eqnarray}
  \frac{Me^{(1)}_n(u)}{Me^{(1)\;\prime}_n(0)}\simeq C\frac{|S(u)-S_*|^{\frac{1}{6}}}{[S'(u)]^{\frac{1}{2}}}\times\nonumber\\
\left\{\ic \Ai\left[\sgn(u_*-u)\left|\frac{3}{2}\left[S(u)-S_*\right]\right|^{2/3}\right]+
\Bi\left[\sgn(u_*-u)\left|\frac{3}{2}\left[S(u)-S_*\right]\right|^{2/3} \right]\right\}
\label{Me_turn_1}
\end{eqnarray}
Consider now the asymptotics at small $u$, i.e. $u\ll u_*$, which also corresponds to $|S(u)-S_*|\to\infty$. The relevant asymptotics for the Airy functions is now:
\begin{equation}
  \label{asympt_Airy+}
  \Ai(x) \sim  \frac{e^{-\frac{2x^{3/2}}{3}}}{2\sqrt{\pi} x^{1/4}},\
  \Bi(x) \sim \frac{e^{\frac{2x^{3/2}}{3}}}{\sqrt{\pi} x^{1/4}},\ 
x\to\infty
\end{equation}
Inserting (\ref{asympt_Airy-}) into (\ref{Me_turn_1}) gives:
\begin{equation}
  \frac{Me^{(1)}_n(u)}{Me^{(1)\;\prime}_n(0)}\simeq\frac{C}{\sqrt{\pi}\left(\frac{3}{2}\right)^{1/6}[S'(u)]^{\frac{1}{2}}}
\left[ \frac{\ic}{2}e^{-(S_*-S)}+ e^{S_*-S}\right] \label{WKBsmallu_naive}
\end{equation}
By definition the left hand side has a unit derivative at $u=0$. Before differentiating the right hand side, 
it is noticed that the regime $u\to 0$ corresponds to $S_*\gg S$, so that, in (\ref{WKBsmallu_naive}) the second term is exponentially larger than the first term. It is then only enough to differentiate this second term within our approximation. This gives:
\begin{equation}
  C=-\sqrt{\pi}\left(\frac{3}{2}\right)^{1/6} \frac{e^{-S_*}}{ S'(0)^{1/2}}\ .\label{valueC}
\end{equation}
Putting (\ref{valueC}) back into (\ref{Me_turn_1}) leads to (\ref{Langer_Men_cross}).

Now we turn to the WKB formulae (\ref{Langer_Me2n_inside}), and (\ref{Langer_Me2n+1_inside}). Eq.~(\ref{WKBsmallu_naive}) was found to be not accurate enough in our numerical checks. Below it is indicated how to circumvene this limitation. 
%We should stress already that all the WKB expressions should agree at large $n$. 

For small $u$, the standard WKB theory claims that one should use for the third radial Mathieu function the formula (\ref{WKB_inside}) with $A_+$, and $A_-$ complex constants. Instead we start from the relation between $Me^{(1)}_n(u)$ and $Ce_n(u)$. From the WKB expansion of the latter solution, see Eq.~(\ref{WKB_ex_in}) with $A=ce_n(0)S'(0)^{1/2}$, we write the third kind Mathieu function  for small $u$ as:
\begin{equation}
  Me^{(1)}_n(u)\simeq ce_n(0) \left[\frac{S'(0)}{S'(u)}\right]^{1/2} \cosh[S(u)]+\ic \frac{A_+ e^{S(u)}+A_- e^{-S(u)} }{S'(u)^{1/2}}\label{WKBsmallu_naive10}\ ,
\end{equation}
where the remaining $A_+$, and $A_-$ are real unknown constants. We choose to cancel the exponentially smaller term as it was mentioned in (\ref{WKBsmallu_naive}). This leads to take $A_+=0$. Differentiating (\ref{WKBsmallu_naive10}) at $u=0$ leads to
\begin{equation}
  A_-=\ic \frac{ Me^{(1)\;\prime}_n(0)}{S'(0)^{1/2}}\ .
\end{equation}
Eventually we have obtained the following small $u$ approximation for the odd third kind radial Mathieu function:
\begin{equation}
  Me^{(1)}_n(u)\simeq ce_n(0)\left[\frac{S'(0)}{S'(u)}\right]^{1/2} \cosh[S(u)]  - Me^{(1)\;\prime}_n(0) \frac{e^{-S(u)} }{\left[S'(0) S'(u)\right]^{1/2}}\ .\label{Me_WKB_smallu}
\end{equation}
Note that Eq.~\ref{Me_WKB_smallu} is very accurate to extract the first or second kind Mathieu functions at small $u$.
One last step consists of using the exact formulae (\ref{C2n}), and (\ref{C2n+1}) to deduce $Me^{(1)\;\prime}_n(0)$ for both parities of $n$. This gives the final results (\ref{Langer_Me2n_inside}), and (\ref{Langer_Me2n+1_inside}), which we used in our numerical method.
\subsection{WKB approximations for $Ne^{(1)}_n(u)$}

The same method can be directly transposed for the odd symmetry class. The requirements to determine uniquely $Ne^{(1)}_n(u)$ are now:
\begin{itemize}
\item it obeys outgoing radiation condition at infinity,
\item its real part is $Se_n(u)$.
\end{itemize}
The same radiation condition leads to the same relation between the two coefficients to be determined. The analogue of (\ref{Me_turn_1}) is:
\begin{eqnarray}
  \frac{Ne^{(1)}_n(u)}{Ne^{(1)}_n(0)}\simeq D\frac{|S(u)-S_*|^{\frac{1}{6}}}{[S'(u)]^{\frac{1}{2}}}\times\nonumber\\
\left\{\ic \Ai\left[\sgn(u_*-u)\left|\frac{3}{2}\left[S(u)-S_*\right]\right|^{2/3}\right]+
\Bi\left[\sgn(u_*-u)\left|\frac{3}{2}\left[S(u)-S_*\right]\right|^{2/3} \right]\right\}\ .
\label{Ne_turn_1}
\end{eqnarray}
The remaining coefficient $D$ will be determined by expanding this formula for $u\ll u_*$. 
Inserting (\ref{asympt_Airy-}) into (\ref{Me_turn_1}) gives:
\begin{equation}
  \frac{Ne^{(1)}_n(u)}{Ne^{(1)}_n(0)}\simeq\frac{D}{\sqrt{\pi}\left(\frac{3}{2}\right)^{1/6}[S'(u)]^{\frac{1}{2}}}
\left[ \frac{\ic}{2}e^{-(S_*-S)}+ e^{S_*-S}\right]\ . \label{WKBsmallu_naive2}
\end{equation}
By definition the left hand side has a unit value at $u=0$. Neglecting the exponentially smaller term, this gives:
\begin{equation}
  D=\sqrt{\pi}\left(\frac{3}{2}\right)^{1/6} e^{-S_*}S'(0)^{1/2}\ .\label{valueD}
\end{equation}
Putting (\ref{valueD}) back into (\ref{Ne_turn_1}) leads to (\ref{Langer_Nen_cross}).
For the formulae (\ref{Langer_Ne2n_inside}), and (\ref{Langer_Ne2n+1_inside}) the same method as for the even symmetry class was followed. We only give here the result, analog to (\ref{Me_WKB_smallu}):
\begin{equation}
  \label{Ne_WKB_smallu}
   Ne^{(1)}_n(u)\simeq se_n'(0)\frac{ \sinh[S(u)]}{\left[S'(0)S'(u)\right]^{1/2}}  + Ne^{(1)}_n(0) \left[\frac{ S'(0)}{ S'(u)}\right]^{1/2}e^{-S(u)}\ .
\end{equation}
Again the exact formulae (\ref{D2n}), and (\ref{D2n+1}) for $Ne^{(1)}_n(0)$ are then used to get (\ref{Langer_Ne2n_inside}), and (\ref{Langer_Ne2n+1_inside}).
\section{Checking the accuracy of our results via an identity}
\label{check_identity}

The formula (\ref{expandH0}) is the very necessary ingredient in order to derive our expansions for the Green functions. This can be also used as a test to estimate more quantitatively the accuracy of our method.

\begin{figure}[!ht]
  \begin{center}
    \includegraphics[width=\textwidth]{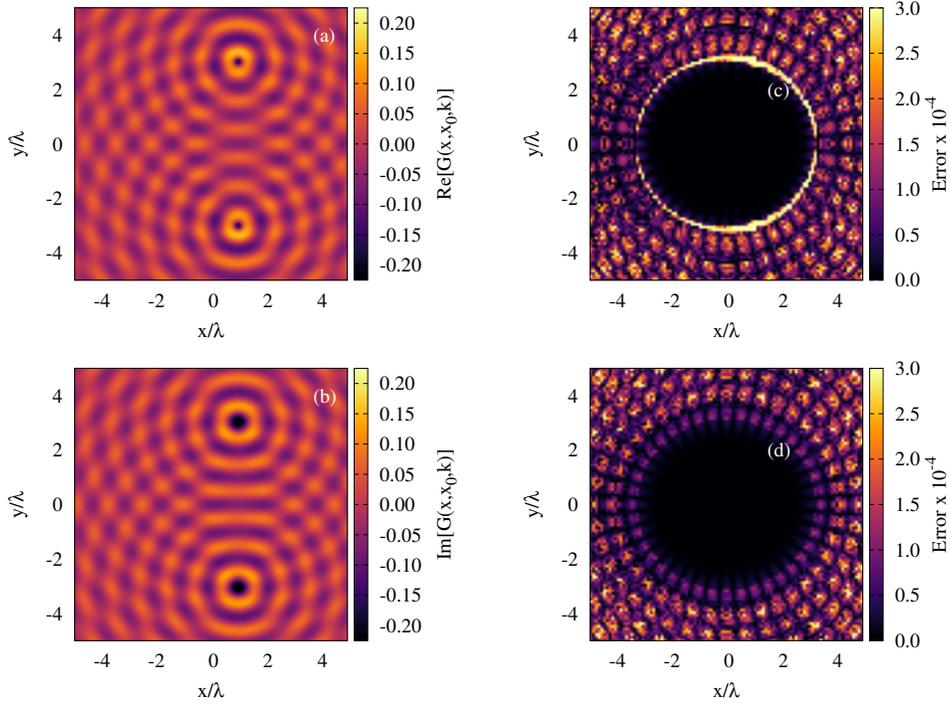}%{Comp_Even.eps}
  \end{center}  
  \caption{Test of Eq.~(\ref{expansionN}) for $(x_0;y_0)=(1,3)$. {(a)}: Real part of the right hand side. (c): Magnified absolute error between the real part of both sides.
(b): Imaginary part of the right hand side. (d): Magnified absolute error between the imaginary part of both sides.}
  \label{testN}
\end{figure}

\begin{figure}[!ht]
  \begin{center}
    \includegraphics[width=\textwidth]{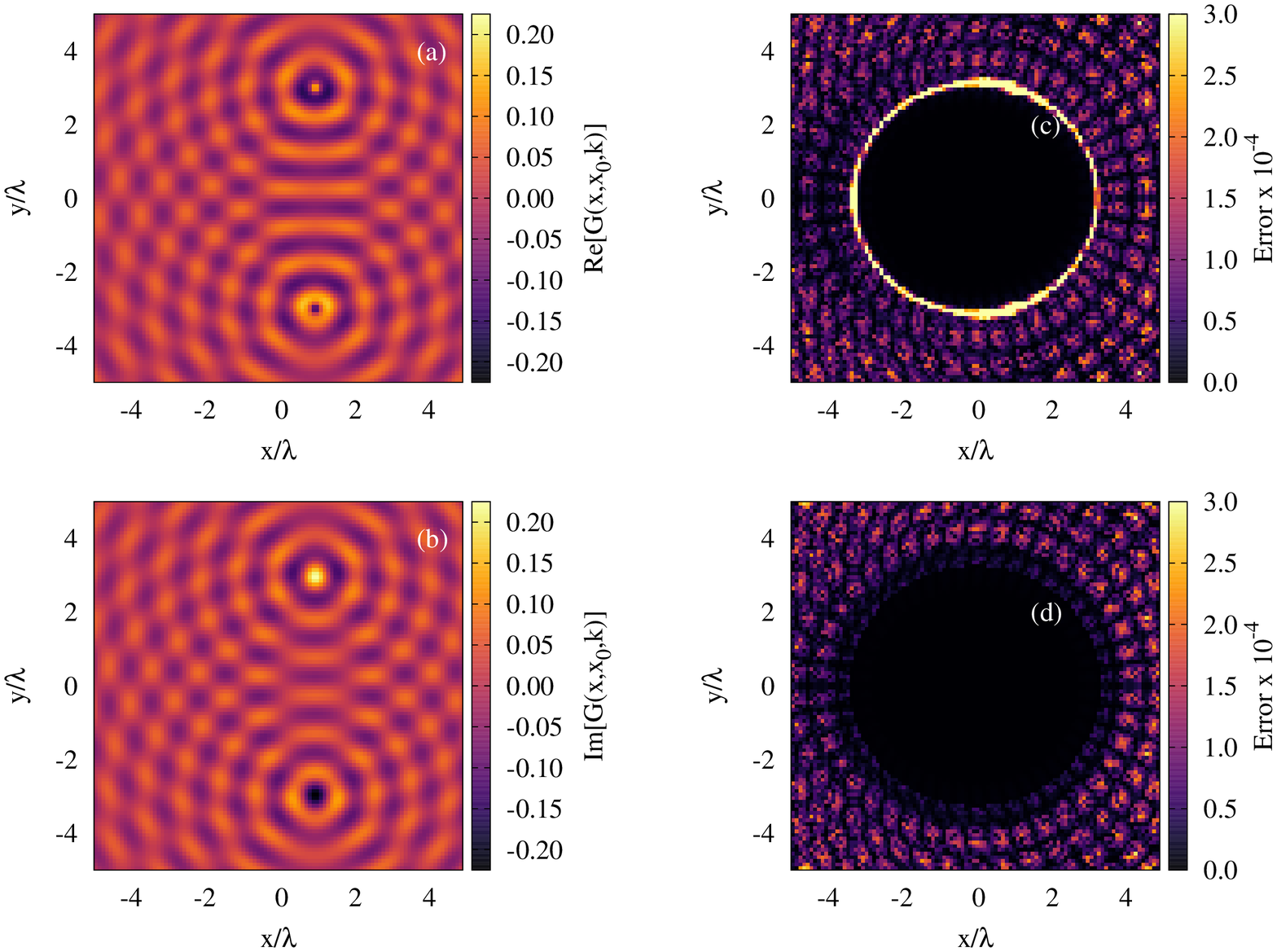}%{Comp_Odd.eps}
  \end{center}  
  \caption{Test of Eq.~(\ref{expansionD}) for $(x_0;y_0)=(1,3)$. {(a)}: Real part of the right hand side. (c): Magnified absolute error between the real part of both sides.
(b): Imaginary part of the right hand side. (d): Magnified absolute error between the imaginary part of both sides.}
  \label{testD}
\end{figure}

From (\ref{expandH0}) it is very easy to deduce the following expansions:
\begin{eqnarray}
  \label{expansionN}
% \hspace{-2cm}
\frac{H_0^{(1)}(k |{\bf x}-{\bf x}_0|)}{4\ic}+\frac{H_0^{(1)}(k |{\bf x}-{\bf x'}_0|)}{4\ic}&=&
\frac{2}{\pi}\displaystyle\sum_{n\ge 0} \frac{Me^{(1)}_n(u_>)Ce_n(u_<)}{Me^{(1)\;\prime}_n(0) ce_n(0) } ce_n(v) ce_n(v_0)\\
  \label{expansionD}
% \hspace{-2cm}
\frac{H_0^{(1)}(k |{\bf x}-{\bf x}_0|)}{4\ic}-\frac{H_0^{(1)}(k |{\bf x}-{\bf x'}_0|)}{4\ic}&=&
-\frac{2}{\pi}\displaystyle\sum_{n>0} \frac{Ne^{(1)}_n(u_>)}{Ne^{(1)}_n(0) }\frac{Se_n(u_<)}{se'_n(0) }se_n(v)se_n(v_0)
\end{eqnarray}
which can be seen as the expansion in a series of Mathieu functions of the Green function for a wave reflected by a wall located at $y=0$. ${\bf x'}_0$ stands for the image of ${\bf x}_0$ under the reflection of the $x-$axis.
This stands to us for a test of both the accuracy of our method and the convergence of the series expansion.
The results for (\ref{expansionN}), resp.~(\ref{expansionD}), are displayed in Fig.~\ref{testN}, resp.~\ref{testD}.

\section{Derivation of the Green function as a series of Mathieu functions}
\label{derive_Green}

First we will need the following fundamental identity, which is proven e.g. in \cite{sips}:
\begin{equation}
  \label{expandH0}
\hspace{-2cm}\frac{H_0^{(1)}(k |{\bf x}-{\bf x}_0|)}{4\ic}=\frac{1}{\pi}
\sum_n \frac{Me^{(1)}_n(u_>) Ce_n(u_<)ce_n(v_0)ce_n(v)}{Me^{(1)\;\prime}_n(0)ce_n(0)} -
\frac{Ne^{(1)}_n(u_>) Se_n(u_<)se_n(v_0)se_n(v)}{Ne^{(1)}_n(0)se'_n(0) }\ ,
\end{equation}

The derivation of the Green function below follows the lines of \cite{strutt}. The solution is expanded in two different series depending on the presence of the source. We assume without loss of generality that the source is the lower half plane, i.e.:
$$-\pi<v_0<0\ .$$
The guessed series comes from the symmetrised part of (\ref{expandH0}), and the required boundary condition.
We choose to consider the following series:
\begin{equation}
\hspace{-2cm}G({\bf x},{\bf x}_0,k)=\left\{
\begin{array}{lc}
\displaystyle\sum_{n\ge 0} \alpha_n^{(+)} Me^{(1)}_n(u)ce_n(v)& 0<v< \pi\\ & \\
\frac{2}{\pi}\displaystyle\sum_{n\ge 0} \frac{Me^{(1)}_n(u_>)Ce_n(u_<)ce_n(v_0) ce_n(v)}{Me^{(1)\;\prime}_n(0) ce_n(0) } +
\displaystyle\sum_{n\ge 0} \alpha_n^{(-)} Me^{(1)}_n(u)ce_n(v)& -\pi < v <0
\end{array}\right.\ . \label{expandG2}
\end{equation}
It is clear that this ansatz is a solution of (\ref{wave_eq}). It also obeys the Neumann boundary condition  along the obstacles. Besides it can be easily shown to obey the matching conditions % (\ref{matchG1}) and (\ref{matchG2}). 
when $|{\bf x}-{\bf x}_0|\to 0$.
Last it obeys the boundary condition at infinity.

The coefficients in the ansatz (\ref{expandG2}) will be fixed by imposing continuity of both the function and its normal derivative along the slit. For simplicity We will assume that the source is not along the slit, i.e.:
$$u_0>0\ .$$
The continuity of the function leads to:
\begin{equation}
  \label{cond1}
  \sum_{n\ge 0} \alpha_n^{(+)} Me^{(1)}_n(0)ce_n(v)=\frac{2}{\pi}\displaystyle\sum_{n\ge 0} \frac{Me^{(1)}_n(u_0)}{Me^{(1)\;\prime}_n(0)}ce_n(v_0) ce_n(-v)+\sum_{n\ge 0} \alpha_n^{(-)} Me^{(1)}_n(0)ce_n(-v)
\end{equation}
For the continuity of the {normal derivative accros the slit}, we will use the following identity:
\begin{equation}
  \label{grady0}
%\frac{\partial G}{\partial y}\Big|_{y=0}
\frac{\partial G}{\partial n}=\frac{2}{a\sin v}\frac{\partial G}{\partial u}\Big|_{u=0} \ .
\end{equation}
Note that the region $y>0$ corresponds to $v>0$ while $y<0$ corresponds to $v<0$.
The continuity of the derivative along $y$ direction leads to:
\begin{equation}
  \label{cond2}
  \sum_{n\ge 0} \alpha_n^{(+)} Me^{(1)\;\prime}_n(0)ce_n(v)=-\sum_{n\ge 0} \alpha_n^{(-)} Me^{(1)\;\prime}_n(0)ce_n(-v)
\end{equation}
It should be stressed that \Cref{cond1,cond2} are valid for $0<v<\pi$. 

Next we will project \Cref{cond1,cond2} along a given $ce_p(v)$. We first use that $ce_n$ is an even function of $v$. Second one has the following orthogonality relations, which can be deduced from (\ref{orthoM}), (\ref{Fourier_ce_even}) and (\ref{Fourier_ce_odd}):
\begin{eqnarray}
  \label{ortho_ce1}
\displaystyle\int_0^\pi ce_n(v)ce_p(v)\ud v = &\frac{\pi}{2}\delta_{n,p}
\end{eqnarray}
Eq.~(\ref{ortho_ce1}) allows to simplify immediately (\ref{cond1}), and (\ref{cond2}):
\begin{eqnarray}
  \label{cond11}
  \alpha_p^{(+)} Me^{(1)}_p(0)&-\alpha_p^{(-)} Me^{(1)}_p(0)=& \frac{2}{\pi}\frac{Me^{(1)}_p(u_0)ce_p(v_0)}{Me^{(1)\;\prime}_p(0)} \\
\label{cond22}
\alpha_p^{(+)} Me^{(1)\;\prime}_p(0)&+\alpha_p^{(-)} Me^{(1)\;\prime}_p(0)=&0
\end{eqnarray}
The determinant of the system is:
$$2Me^{(1)}_p(0)Me^{(1)\;\prime}_p(0)\ ,$$
hence never vanishes so the solution is unique as expected. The inversion of the system leads to:
\begin{eqnarray}
  \label{aplus}
  \alpha_p^{(+)}&=&\frac{1}{\pi}\frac{Me^{(1)}_p(u_0)ce_p(v_0)}{Me^{(1)}_p(0)Me^{(1)\;\prime}_p(0)}\\
  \label{aminus}
  \alpha_p^{(-)}&=&-\frac{1}{\pi}\frac{Me^{(1)}_p(u_0)ce_p(v_0)}{Me^{(1)}_p(0)Me^{(1)\;\prime}_p(0)}
\end{eqnarray}
Putting the expressions (\ref{aplus}), and (\ref{aminus}) back into (\ref{expandG2}) gives the solution of the problem, see Eq.~(\ref{GsingleslitN_v2}).
\section{Green function for the scattering by a strip}
\label{Strip}

The method exposed in the previous Section in order to solve the problem by a single slit can be directly transposed to the scattering by a strip. For the sake of completeness, we give below the corresponding formula for the Green function for both Neumann and Dirichlet boundary condition. We also show pictures of the Green function obtained via our numerical approach.

\subsection{Neumann boundary condition}

The Green function for a wave scattered by a strip or a ribbon can be also obtained as series of Mathieu function. The important assumption is to see the strip with a shrunk elliptical obstacle so as to start with the following ansatz:
\begin{equation}
  \label{Gstriptrial}
  G({\bf x},{\bf x}_0,k)=\frac{H_0^{(1)}(k |{\bf x}-{\bf x}_0|)}{4\ic}
+\displaystyle\sum_{n\ge 0} \alpha^{(+)}_n Me^{(1)}_n(u) ce_n(v) + \displaystyle\sum_{n> 0} \alpha^{(-)}_n Ne^{(1)}_n(u) se_n(v) 
\end{equation}
The method to find the unknown coefficients $\alpha^{(+)}_n$ and $\alpha^{(-)}_n$, follows the same lines as in the previous Section. The trial function (\ref{Gstriptrial}) has been chosen to fulfill outgoing boundary condition at infinity and the matching conditions when $|{\bf x}-{\bf x}_0|$ goes to $0$. Decomposing the first term into series of Mathieu functions using (\ref{expandH0}), the boundary condition along the strip, i.e. at $u=0$, is explicited. Projecting this condition along $ce_p(v)$ and using the orthogonality relation (\ref{ortho_ce1}) leads to $\alpha^{(+)}_p$, while the analog projection along $se_p(v)$ leads to $\alpha^{(-)}_p$.
The result can be written in the following way:
\begin{eqnarray}
  \label{GstripN}
  G({\bf x},{\bf x}_0,k)=\frac{H_0^{(1)}(k |{\bf x}-{\bf x}_0|)}{8\ic}+\frac{H_0^{(1)}(k |{\bf x}-{\bf x'}_0|)}{8\ic}-\nonumber\\
\frac{1}{\pi}\displaystyle\sum_{n> 0} \left[\frac{Ne^{(1)}_n(u_>)}{Ne_n^{(1)}(0) }\frac{Se_n(u_<)}{se'_n(0) } -
\frac{Ne^{(1)}_n(u_0)}{Ne^{(1)}_n(0)} \frac{Ne^{(1)}_n(u)}{Ne^{(1)\;\prime}_n(0)} \right] se_n(v_0) se_n(v)\ .
\end{eqnarray}

\begin{figure}[!ht]
  \begin{center}
    \includegraphics[width=\textwidth]{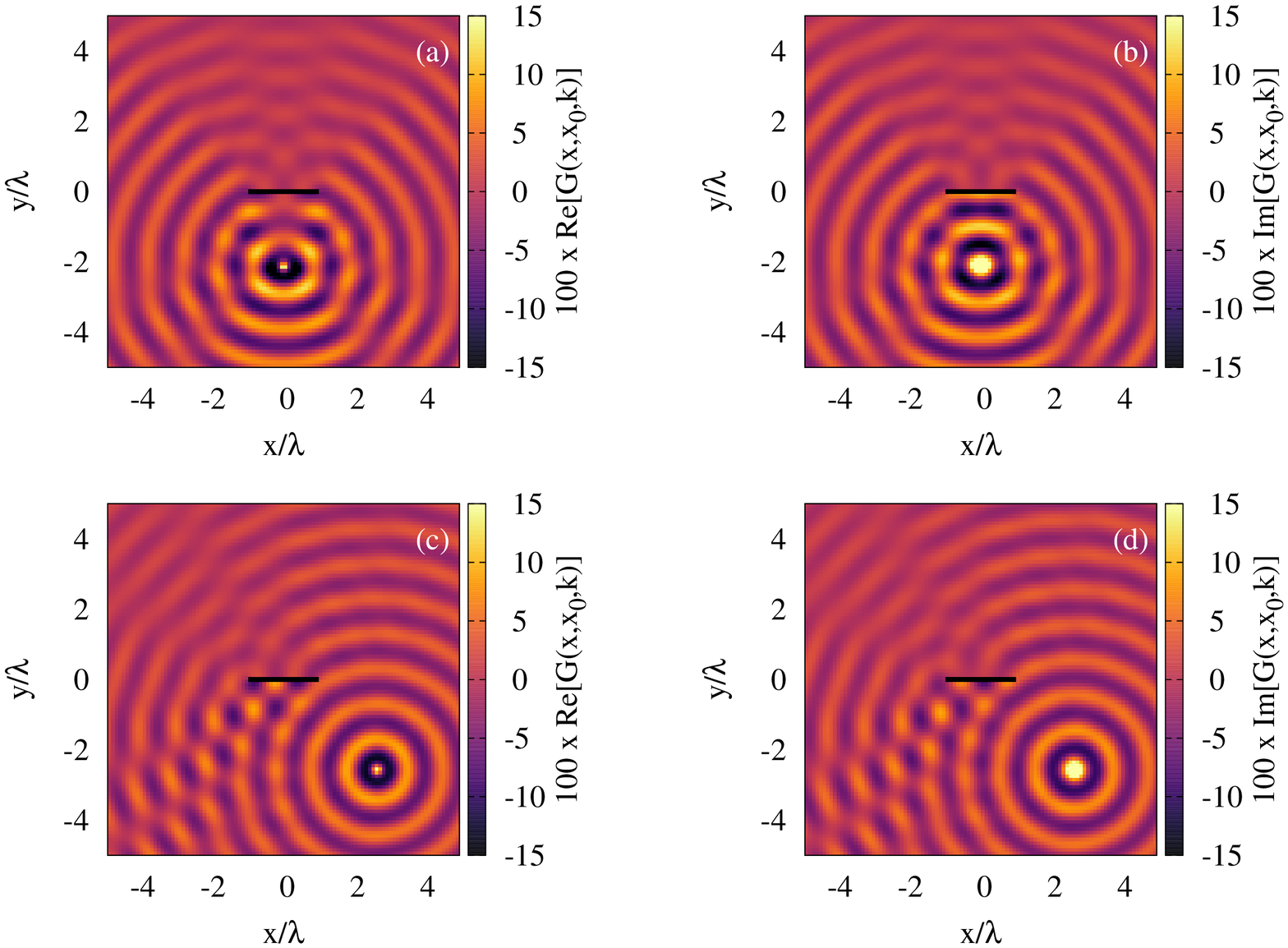}
  \end{center}  
  \caption{Examples of Green function for the scattering by a single strip with Neumann boundary condition with $a/\lambda = 2$, i.e. $\theta = \pi^2$. (a) and (b): diffraction for a source at $(x_0,y_0)\simeq(0,-2.129)$, real and imaginary part respectively. {(c) and (d)}: diffraction for a source at $(x_0,y_0)\simeq(2.660,-2.565)$, real and imaginary part respectively.}
  \label{GreenstripN}
\end{figure}

The result of our numerical computation is illustrated in Fig.~\ref{GreenstripN}.

\subsection{Dirichlet boundary condition}

An identical reasoning as in the previous paragraph can be followed for the case of Dirichlet boundary condition. The Green function is then:
\begin{eqnarray}
  \label{GstripN}
  G({\bf x},{\bf x}_0,k)=\frac{H_0^{(1)}(k |{\bf x}-{\bf x}_0|)}{8\ic}-\frac{H_0^{(1)}(k |{\bf x}-{\bf x'}_0|)}{8\ic}+\nonumber\\
\frac{1}{\pi}\displaystyle\sum_{n\ge 0} \left[\frac{Me^{(1)}_n(u_>)}{Me_n^{(1)\;\prime}(0) }\frac{Ce_n(u_<)}{ce_n(0) } -
\frac{Me^{(1)}_n(u_0)}{Me^{(1)\;\prime}_n(0)} \frac{Me^{(1)}_n(u)}{Me^{(1)}_n(0)} \right] ce_n(v_0) ce_n(v)\ .
\end{eqnarray}

\begin{figure}[!ht]
  \begin{center}
    \includegraphics[width=\textwidth]{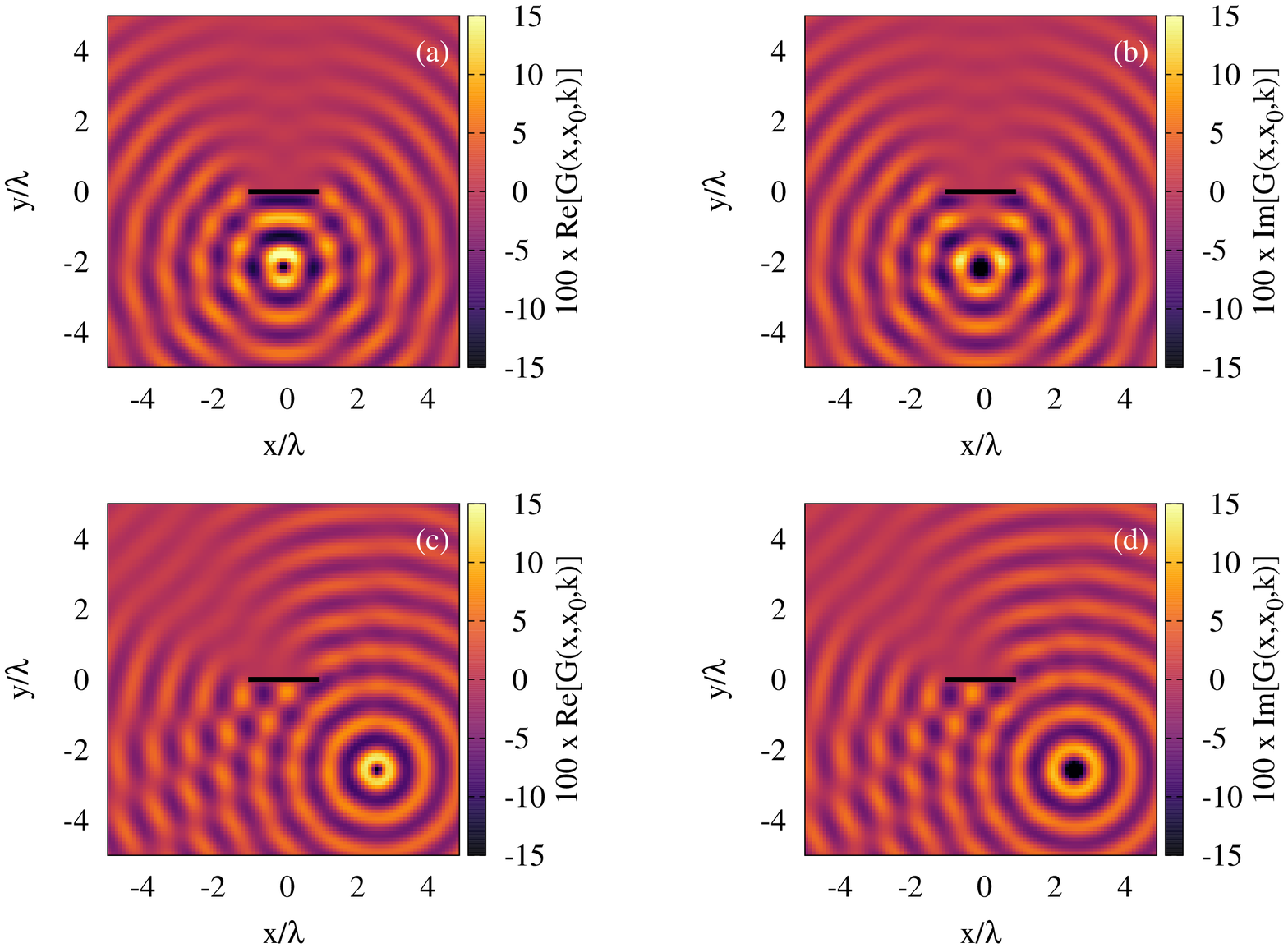}
  \end{center}  
  \caption{Examples of Green function for the scattering by a single strip with Dirichlet boundary condition with $a/\lambda = 2$, i.e. $\theta = \pi^2$. (a) and (b): diffraction for a source at $(x_0,y_0)\simeq(0,-2.129)$, real and imaginary part respectively. {(c) and (d)}: diffraction for a source at $(x_0,y_0)\simeq(2.660,-2.565)$, real and imaginary part respectively.}
  \label{GreenstripD}
\end{figure}

The result of our numerical computation is illustrated in Fig.~\ref{GreenstripD}.

%% References
%%
%% Following citation commands can be used in the body text:
%% Usage of \cite is as follows:
%%   \cite{key}         ==>>  [#]
%%   \cite[chap. 2]{key} ==>> [#, chap. 2]
%%

%% References with bibTeX database:

% \bibliographystyle{elsarticle-num}
% \bibliography{<your-bib-database>}

%% Authors are advised to submit their bibtex database files. They are
%% requested to list a bibtex style file in the manuscript if they do
%% not want to use elsarticle-num.bst.

%% References without bibTeX database:

\end{document}